\documentclass[twocolumn,aps,prc,superscriptaddress,showpacs]{revtex4}

\usepackage{multirow}
\usepackage{epsfig}
\usepackage{amsmath}

\usepackage[normalem]{ulem}  
\usepackage[dvips]{color} 
\newcommand{\com}[1]{{\color[rgb]{0,0,1}{#1}}}

\renewcommand\sout{\bgroup \color{red} \ULdepth=-.5ex \ULset}

\newcommand{\qbar}{\bar{q}}
\newcommand{\sbar}{\bar{s}}
\newcommand{\cbar}{\bar{c}}
\newcommand{\bbar}{\bar{b}}
\newcommand{\VEV}[1]{\left\langle{#1}\right\rangle}
\newcommand{\MeV}{\mathrm{MeV}}

\newcommand{\Ex}[2]{\ifmmode{#1\times10^{#2}}\else{$#1\times10^{#2}$}\fi}
\newcommand{\Exm}[2]{#1\times10^{#2}}
\newcommand{\Psfig}[2]{\includegraphics[width=#1]{#2}}

\begin{document}
\title{Studying Exotic Hadrons in Heavy Ion Collisions}

\author{Sungtae Cho}\affiliation{Institute of Physics and Applied Physics, Yonsei
University, Seoul 120-749, Korea}
\author{Takenori Furumoto}\affiliation{Yukawa Institute for Theoretical Physics, Kyoto University, Kyoto 606-8502, Japan}\affiliation{RIKEN Nishina Center, Hirosawa 2-1, Wako, Saitama 351-0198, Japan}
\author{Tetsuo Hyodo}\affiliation{Department of Physics, Tokyo Institute of Technology, Meguro 152-8551, Japan}
\author{Daisuke Jido}\affiliation{Yukawa Institute for Theoretical Physics, Kyoto University, Kyoto 606-8502, Japan}
\author{Che Ming Ko}\affiliation{Cyclotron Institute and  Department of Physics and Astronomy, Texas A\&M
University, College Station, Texas 77843, U.S.A.}
\author{Su Houng~Lee}\affiliation{Institute of Physics and Applied Physics, Yonsei
University, Seoul 120-749, Korea}\affiliation{Yukawa Institute for
Theoretical Physics, Kyoto University, Kyoto 606-8502, Japan}
\author{Marina Nielsen}\affiliation{Instituto de F\'{\i}sica, Universidade de S\~{a}o Paulo,
C.P. 66318, 05389-970 S\~{a}o Paulo, SP, Brazil}
\author{Akira Ohnishi}\affiliation{Yukawa Institute for Theoretical Physics, Kyoto University, Kyoto 606-8502, Japan}
\author{Takayasu Sekihara}\affiliation{Yukawa Institute for Theoretical Physics, Kyoto University, Kyoto 606-8502, Japan}\affiliation{Department of Physics, Graduate School of Science, Kyoto University, Kyoto 606-8502, Japan}
\author{Shigehiro Yasui}\affiliation{Institute of Particle and Nuclear Studies, High Energy Accelerator Research Organization (KEK), 1-1, Oho, Ibaraki 305-0801, Japan}
\author{Koichi Yazaki}\affiliation{Yukawa Institute for Theoretical Physics, Kyoto University, Kyoto 606-8502, Japan}\affiliation{Hashimoto Mathematical Physics Lab., Nishina Center, RIKEN, 2-1, Hirosawa, Wako, Saitama 351-0198, Japan}

\collaboration{ExHIC Collaboration}\noaffiliation

\date{\today}
\begin{abstract}
We investigate the possibilities of using measurements in present
and future experiments on heavy ion collisions to answer some
longstanding problems in hadronic physics, namely identifying
hadronic molecular states and exotic hadrons with multiquark
components. The yields of a selected set of exotic hadron
candidates in relativistic heavy ion collisions are discussed in
the coalescence model in comparison with the statistical model. We
find that the yield of a hadron is typically an order of magnitude
smaller when it is a compact multiquark state, compared to that of
an excited hadronic state with normal quark numbers. We also find
that some loosely bound hadronic molecules are formed more
abundantly than the statistical model prediction by a factor of
two or more. Moreover, due to the significant numbers of charm and
bottom quarks produced at RHIC and even larger numbers expected at
LHC, some of the proposed heavy exotic hadrons could be produced
with sufficient abundance for detection, making it possible to
study these new exotic hadrons in heavy ion collisions.
\end{abstract}

\pacs{14.40.Gx,11.55.Hx,24.85.+p}

\maketitle

\section{Introduction}

Experiments at the Relativistic Heavy Ion Collider (RHIC) during
the past decade have shown that the properties of the quark-gluon
plasma (QGP) formed in heavy ion collisions are far more
intriguing than originally conceived~\cite{qm09}. Instead of
weakly interacting, the quark-gluon plasma was found to be a
strongly coupled system with so small a shear viscosity that it
behaves like an ideal fluid. The study of the QGP is expected to
remain an active field of research in the future because of the
proposed upgrade of RHIC and new experimental possibilities at the
Large Hadron Collider (LHC). The physics of QGP is related to a
wide range of other fields, such as the early universe,
nonequilibrium statistical physics, string theory by AdS/CFT
correspondence, and so on.

During the same time, there have also been exciting developments
in the spectroscopy of heavy hadrons, starting with the discovery
of the $D_{sJ}(2317)$ by the BABAR
Collaboration~\cite{Aubert:2003fg}, whose mass could not be
explained by the simple quark model, and the $X(3872)$ by the
Belle Collaboration~\cite{Choi:2003ue}, whose mass and decay
channel strongly support a nontrivial fraction of $\bar{D}D^*$ and
$D\bar{D}^*$ components in its wave function. The Belle
collaboration also reported the finding of the Z$^{+}$(4430) in
the $\psi'\pi^+$ spectrum \cite{Belle:Z4430}. Because the
Z$^{+}$(4430) is a charged state, it cannot be a simple $c\bar{c}$
state. If confirmed, this would be the first evidence for the
existence of an exotic hadron that is composed of two quarks and
two antiquarks like $c\bar{c}u\bar{d}$~\cite{Marina-review}.

The question of whether multiquark hadrons exist is an old problem
in the light hadron sector that began with attempts to understand
the inverted mass spectrum of the scalar nonet ($a_{0}(980)$,
$f_{0}(980)$, and so on) in the tetraquark
picture~\cite{Jaffe76-1,Jaffe76-2}.  Also, the exotic H dibaryon
was proposed on the basis of the color-spin
interaction~\cite{Jaffe76}, and it has been sought for in various
experiments for a long time without success.  On the other hand,
the $\Lambda(1405)$ baryon resonance was considered as a
$\bar{K}N$ quasi-bound state in the $\bar{K}N-\pi\Sigma$
coupled-channel analysis~\cite{Dalitz:1960du}. With further
development of the coupled-channel approach~\cite{Kaiser}, it has
been realized that the $\Lambda(1405)$ is the most obvious and
uncontroversial candidate for a hadronic molecule, whose wave
function is composed dominantly of a $\bar{K}N$ bound state mixed
with a small $\pi \Sigma$ resonant state~\cite{Hyodo:2011ur}.  If
such a configuration for the $\Lambda(1405)$ is confirmed in
experimental measurements, it will be the first evidence for a
molecular hadronic state. Many new multiquark states such as
$\bar{K}KN$, $\bar{K}NN$, $(\Omega\Omega)_{0}$, and {\it etc.}
have also been predicted.

While the Z$^{+}$(4430) could be a first explicitly exotic hadron
found to date, it will be a milestone in hadronic spectroscopy if
other flavor exotic hadrons are found, such as the controversial
pentaquark $\Theta^+(udud\bar s)$ first reported in experiments on
photonuclear reactions~\cite{Nakano:2003qx}. Also, several other
flavor exotic molecular and compact multiquark hadrons were
previously predicted, based, respectively, on the meson-exchange
and color-spin interactions. More recently, a simple diquark model
based on the color-spin interaction~\cite{Lee07,Lee09} was shown
to naturally explain the likely existence of flavor exotic
multiquark hadrons consisting of heavy spectator quarks, such as
the $T^1_{cc}(ud\bar{c}\bar{c})$, $T^{0}_{cb}(ud\bar{c}\bar{b})$,
and $\Theta_{cs}(udus\bar{c})$ that were predicted before, and the
newly predicted dibaryon $H^{++}_c(udusuc)$. Furthermore, the
stable molecular bound states $\bar{D}N$ and $\bar{D}NN$ in the
charmed sector and the $BN$ and $BNN$ in the bottom sector have
been predicted to exist as a result of the long-range
pion-exchange potential~\cite{Yasui:2009bz,Yamaguchi:2011xb}.

To gain insights into all these proposals and
questions~\cite{Chen:2003tn,Chen07,Lee07}, we have proposed in
recent publications~\cite{Cho:2010db, Ohnishi:2011nq} a new
approach of studying exotic hadrons in heavy ion collisions at
ultrarelativistic energies. There are several merits for this
approach compared to the search for exotic hadrons in elementary
particle reactions that have been pursued so far. First, an
appreciable number of heavy quarks are expected to be produced in
these collisions, reaching as large as 20 $\bar{c}c$ pairs per
unit rapidity in Pb+Pb collisions at the Large Hadron Collider
(LHC)~\cite{Zhang08}. Second, through vertex reconstruction of
weakly decaying particles, one could substantially reduce the
backgrounds in the detection, making the identification of weakly
decaying exotics possible. Finally, because of the large volume of
quark and hadronic matters formed in these collisions and the new
paradigm of hadronization through
coalescence~\cite{Hwa03,Gre03,Fri03}, various exotic hadrons could
be formed from the recombinations of quarks. We have therefore
investigated in Refs.~\cite{Cho:2010db,Ohnishi:2011nq} the
possibility of using measurements in present and future
experiments on heavy ion collisions to answer the longstanding
problems in hadronic physics of identifying and examining hadronic
molecular states and exotic hadrons that consist of multiquarks.

In the present paper, we extend the discussions in
Ref.~\cite{Cho:2010db} to include all exotic
hadrons that have been proposed so far. These hadrons are
classified as exotic mesons, exotic baryons, and exotic dibaryons,
and for each exotic hadron we give a brief summary of its
properties and the status on the latest researches.
Moreover, we collect all necessary references so that
the present paper can be used as a general guide to the literatures
on exotic hadrons. Using these information, we then evaluate their yields
in heavy ion collisions based on both the statistical and the coalescence model.
In particular, we present a detailed description of the calculations
carried out in Ref.~\cite{Cho:2010db} with an emphasis on the
expected yields of exotic hadrons in heavy ion collisions at LHC.

The statistical model is based on the assumption that hadrons
produced in relativistic heavy ion collisions are in thermal and
chemical equilibrium at temperatures that are close to that for
the QGP to hadronic matter phase transition, and it has been known
to describe very well the relative yields of normal
hadrons~\cite{Andronic06}. The coalescence model, on the other
hand, describes hadron production through the coalescence or
recombination of particles~\cite{Hwa03,Gre03,Fri03}. The model has
successfully explained observed enhancement in the production of
midrapidity baryons in the intermediate transverse momentum
region~\cite{Adcox:2001mf,Abelev:2007ra}, the quark number scaling
of the elliptic flow of identified
hadrons~\cite{Adler:2003kt,Sorensen:2003wi}, and the yield of
antihypertritons recently discovered in heavy ion collisions at
RHIC~\cite{Abelev:2010rv}. Furthermore, the coalescence model that
takes into account the internal structure of hadrons has been
shown in Ref.~\cite{KanadaEn'yo:2006zk} to be able to describe the
observed suppression of the $\Lambda(1520)$ yield in heavy ion
collisions at RHIC compared to the prediction of the statistical
model~\cite{abelev}. As in Ref.~\cite{Cho:2010db}, we fix the
parameters in the coalescence model by reproducing the yields of
normal hadrons in the statistical model~\cite{Andronic06}. We then
apply these two models to calculate the yields of exotic hadrons.
As reference particles for the comparison with exotic hadrons, we
consider, in particular, normal hadrons such as the strange
$\Lambda(1115)$ and $\Lambda(1520)$, the charmed
$\Lambda_c(2286)$, and the bottom $\Lambda_b(5620)$. Our results
show that the yields of proposed exotic mesons, baryons, and
dibaryons are large enough for carrying out realistic
measurements. We further discuss their most probable weak decay
channels that can be observed in experiments.

Our study shows that results from the coalescence model are
sensitive to the inner structure of hadrons, such as the angular
momentum, quark numbers and so on. This is different for the
statistical model, because the yields of the hadrons in this model
are determined mainly by their masses. To discriminate the
different pictures for exotic hadrons, we can thus compare the
results from the coalescence model with those from the statistical
model. We find that the relative yields of light exotic hadrons in
the coalescence model are very different from those in the
statistical model, and this makes it possible to experimentally
discriminate among the different pictures for their structures,
such as multiquarks versus hadronic molecules.

This paper is organized as follows. In Sec.~\ref{coal}, we
describe briefly the schematic model used for describing the time
evolution of the hot dense matter formed in relativistic heavy ion
collisions. We then explain the statistical model and the
coalescence model used in the present study. The properties of the
exotic hadrons included in this work and their predicted yields
are given in Secs.~\ref{properties} and \ref{results},
respectively. This is followed by discussions in
Sec.~\ref{discussion} and conclusions in Sec.~\ref{sec:summary}.
In Appendix A, we derive the Wigner function of hadrons whose
structures are described by the $d$-wave and the corresponding
coalescence factor. We further give in Appendix B the coalescence
factors for the general $l$-wave harmonic oscillator wave
functions.

\section{Models for hadron production}\label{coal}

\subsection{Heavy ion collision dynamics}

For the dynamics of central relativistic heavy ion collisions, we
use the schematic model of Ref.~\cite{Chen:2003tn} based on
isentropic boost invariant longitudinal and accelerated transverse
expansions. In this model, both the initial quark-gluon plasma and
final hadronic matter are treated as noninteracting free gas, and
the transition between these two phases of matter is taken to be
first-order. The time evolution of the temperature and volume of
the system is determined by the entropy conservation. In
Table~\ref{parametersA}, we tabulate the values of the critical or
hadronization temperature $T_C$ and volume $V_C$ at the beginning
of the quark-gluon to hadronic matter phase transition, the volume
$V_H$ at the end of the mixed phase or hadronization, and the
kinetic or thermal freeze-out temperature $T_F$ and volume $V_F$
for both central Au+Au collisions at $\sqrt{s_{NN}}=200$ GeV at
RHIC and central Pb+Pb collisions at $\sqrt{s_{NN}}=5$ TeV at LHC.
Also given in Table~\ref{parametersA} is the abundance of various
quarks at $T_C$.

\begin{table}[htdp]
\caption{Quark numbers at hadronization temperature $T_C$ and
volume $V_C$, the volume $V_H$ at the end of hadronization, and
the thermal freeze-out temperature $T_F$ and volume $V_F$ in
central heavy ion collisions at RHIC and LHC.} \label{parametersA}
\begin{tabular}{cccc}
\hline
\hline
 & RHIC & LHC \nonumber \\ \hline
$N_u=N_d$ & 245 &662 \nonumber \\
$N_s=N_{\bar{s}}$ & 150 & 405 \nonumber \\
$N_c=N_{\bar{c}}$ & 3 & 20 \nonumber \\
$N_b=N_{\bar{b}}$ & 0.02 & 0.8 \nonumber \\
$V_C$ & 1000 fm$^3$ & 2700 fm$^3$ \nonumber \\
$T_C=T_H$ & 175 MeV & 175 MeV \nonumber \\
$V_H$ & 1908 fm$^3$ &  5152 fm$^3$  \nonumber \\
$\mu_B$ & 20 MeV & 0 MeV \\
$\mu_s$ & 10 MeV & 0 MeV \\
\hline
$V_F$ & 11322 fm$^3$ & 30569 fm$^3$  \nonumber \\
$T_F$ & 125 MeV  & 125 MeV \nonumber \\
\hline\hline
\end{tabular}
\end{table}

As described in next subsections, hadron production
in the statistical model occurs in the volume $V_H$
at the temperature $T_H$, which is assumed to be the same as
$T_{C}$, while in the quark coalescence model, the production of
both ordinary and multiquark hadrons occurs at the temperature $T_C$ when the
volume of the QGP is $V_C$. For the production of hadronic molecular states
from the coalescence of hadrons, it takes place, on the other hand,
in the volume $V_F$ at the kinetic freeze-out temperature $T_F$.

\subsection{The statistical model}

The statistical model has been shown to describe very well the
relative yields of normal hadrons in relativistic heavy ion
collisions. In this model, the number of produced hadrons of a
given type $h$ is given by~\cite{Andronic06}
\begin{align}
\label{Eq:Stat}
N_h^\mathrm{stat} = & V_H \frac{g_h}{2 \pi^2}
\int_0^\infty \frac{p^2 dp}{\gamma_h^{-1}e^{E_h/T_H} \pm 1} \\
\label{Eq:StatSimple}
\approx & \frac{\gamma_hg_hV_H}{2\pi^2}m_h^2T_HK_2(m_h/T_H)\\
\label{Eq:StatNR} \approx &
\gamma_hg_hV_H\left(\frac{m_hT_H}{2\pi}\right)^{3/2}e^{-m_i/T_H},
\end{align}
In the above equations, $g_h$ is the degeneracy of the hadron and
$\gamma_h$ is the fugacity, and $V_H$ and $T_H$ are, respectively,
the volume and temperature of the source when statistical
production occurs.

Since strangeness is known to reach approximate chemical
equilibrium in heavy ion collisions at RHIC due to the short
equilibration time in the quark-gluon plasma and the net
strangeness of the QGP is zero, the strange chemical potential is
small and is taken to be $\mu_s=10 $ MeV. Its value decreases with
increasing collision energy and is assumed to be 0 MeV in heavy
ion collisions at LHC. For charm and bottom quarks, they are
produced from initial hard scattering and their numbers are much
larger than those expected from a chemically equilibrated
quark-gluon plasma. As a result, we obtain the fugacity
$\gamma_h>1$ for both charmed and bottom hadrons. In terms of the
fugacities $\gamma_c$ and $\gamma_b$ of charm and bottom quarks,
the fugacities of charmed and bottom hadrons are products of
$\gamma_c^n$ and $\gamma_b^m$ where $n$ and $m$ are, respectively,
the charm and bottom quark numbers in these hadrons. Therefore,
the fugacity of hadron species $h$ can be written generally as
\begin{align}
\gamma_h= \gamma_c^{n_c+n_{\bar{c}}}
\gamma_b^{n_b+n_{\bar{b}}} e^{(\mu_B B + \mu_s S)/T_H}\ ,
\end{align}
where $B$, $S$, $n_c (n_{\bar{c}})$, $n_b (n_{\bar{b}})$ are the
baryon number, strangeness, (anti-)charm quark number, and
(anti-)bottom quark number of the hadron, respectively.

For the charm and bottom fugacities $\gamma_c$ and $\gamma_b$,
they can be determined by requiring that the total yield of
charmed or bottom hadrons estimated in the statistical model is
the same as the expected total charm ($N_c$) or bottom ($N_b$)
quark number from initial hard nucleon-nucleon scattering. Using
the values $N_c=3$ and $N_b=0.02$ for heavy ion collisions at
RHIC, we obtain $\gamma_c=6.40$ and $\gamma_b=2.2\times 10^6$
according to the following calculations:
\begin{eqnarray}
N_c &= & N_{D}+N_{D^*}+\frac12\left(N_{D_s}+N_{\bar{D}_s}\right)
    +\frac12\left(N_{\Lambda_c}+N_{\bar{\Lambda}_c}\right)
     \nonumber \\
&=&
1.04+1.53+\frac{0.33+0.29}{2}+\frac{0.14+0.11}{2}=3,\label{Eq:number_c}
\end{eqnarray}
\begin{eqnarray}
N_b &= & N_{\bar{B}}+N_{\bar{B}^*}
    +\frac12\left(N_{\bar{B}_s}+N_{B_s}\right)
    +\frac12\left(N_{\Lambda_b}+N_{\bar{\Lambda}_b}\right)
     \nonumber \\
&=& \Exm{5.3}{-3}+\Ex{1.23}{-2}
\nonumber\\
&+&\frac{1.7+1.5}{2}\times10^{-3}
+\frac{9.2+7.3}{2}\times10^{-4}=0.02.
\label{Eq:number_b}
\end{eqnarray}
In the above, we have used the average yield of heavy strange and
antistrange mesons as well as that of heavy baryons and antibryons
to remove the effect of baryon and strangeness chemical
potentials. A similar analysis for heavy ion collisions at LHC
based on the charm and bottom quark numbers $N_c=20$ and $N_b=0.8$
(see Table~\ref{parametersA}) gives the charm and bottom
fugacities $\gamma_c=15.8$ and $\gamma_b=3.3\times 10^7$. These
values together with those for RHIC are given in
Table~\ref{parametersB}.

\begin{table}[htdp]
\caption{Fugacities for $c$ and $b$ quarks, and hadron numbers
from the statistical model at
thermal freeze-out temperature $T_F$ and volume $V_F$ in central
heavy ion collisions at RHIC and LHC, including contributions from
resonance decays shown in the fourth column. }\label{parametersB}
\begin{tabular}{cccc}
\hline
\hline
 & RHIC & LHC \nonumber \\ \hline
$\gamma_c$      & 6.40      & 15.8 \\
$\gamma_b$      & \Ex{2.2}{6}   & \Ex{3.3}{7} \\
\hline
$N_{K}$                 & 142           &   363   & $K, K^*$ \\
$N_{\bar{K}}$           & 127           & 363   & $\bar{K}, \bar{K}^*$ \\
$N_D=N_{\bar{D}}$       & 1.0           & 6.9     \\
$N_{D^*}=N_{\bar{D}^*}$ & 1.5           & 10      \\
$N_{D_1}$               & 0.19          & 1.3     \\
$N_B=N_{\bar{B}}$       & \Ex{5.3}{-3}  & 0.21    \\
$N_{B^*}=N_{\bar{B}^*}$ & \Ex{1.2}{-2}  & 0.49    \\
\hline
$N_{N}$                 & 62            & 150   & $N, \Delta$ \nonumber \\
$N_\Xi$             & 4.7           & 13    & \\
$N_\Omega$              & 0.81          & 2.3     \\
$N_{\Xi_c}$         & 0.10          & 0.65  & \\
\hline\hline
\end{tabular}
\end{table}

In Table~\ref{parametersB}, we also show the yield $N_{K}$
($N_{\bar{K}}$) of $K$ ($\bar{K}$) mesons given as a sum of the
directly produced $K$ ($\bar{K}$) and those from the strong decay
of $K^{*}$ ($\bar{K}^{*}$) after their freeze-out. Similarly, the
yield of nucleons $N_{N}$ includes both directly produced $N$ and
those from the strong decay of $\Delta$. These results are
obtained by using the Fermi-Dirac and Bose-Einstein distributions
in Eq.~(\ref{Eq:Stat}).  We note that approximating these
distributions by the Boltzmann distribution as given in
Eq.~(\ref{Eq:StatSimple}) does not introduce a large error (at
most 1 \% for exotic hadrons), while the non-relativistic
approximation used in Eq.~(\ref{Eq:StatNR}) leads to an error of
about 30-40\%.

\subsection{The coalescence model}
\label{subsec:2C}

The coalescence model for particle production in nuclear reactions
is based on the sudden approximation by calculating the overlap of
the density matrix of the constituents in an emission source with
the Wigner function of the produced particle~\cite{Sat81}. The
model has been extensively used to study light nuclei production
in nuclear reactions~\cite{Chen03} as well as hadron production
from the quark-gluon plasma produced in relativistic heavy ion
collisions~\cite{Hwa03,Gre03,Fri03,Che06}. In the coalescence
model, the number of hadrons of certain type $h$ produced from the
coalescence of $n$ constituents is given by~\cite{Gre03}
\begin{eqnarray}
N_h^\mathrm{coal}& =& g_h \int \left[\prod_{i=1}^n \frac{1}{g_i}\frac{p_i \cdot d \sigma_i}{(2 \pi)^3} \frac{\mathrm{d}^3 {\bf p}_i}{E_i} f(x_i,p_i)\right]  \nonumber \\ &&  \times  f^W(x_1,\cdots, x_n:p_1,\cdots,p_n).
\end{eqnarray}
In the above equation, $g_h$ is again the degeneracy of the hadron
whereas $g_i$ is that of its $i$th constituent and d$\sigma_i$
denotes an element of a space-like hypersurface. The function
$f(x_i,p_i)$ is the covariant phase-space distribution function of
the constituents in the emission source, and it is normalized to
their number, i.e.,
\begin{equation}
\int p_i\cdot d\sigma_i\frac{\mathrm{d}^3{\bf p}_i}{(2\pi)^3E_i}f(x_i,p_i)=N_i,
\end{equation}
and the function $f^W(x_1 ...x_n:p_1 ...p_n)$ is the Wigner
function of the produced hadron and is defined by
\begin{align}
&f^W(x_1,\cdots,x_n:p_1,\cdots,p_n)
\nonumber\\
=&\int\prod_{i=1}^n dy_i e^{ip_iy_i} \psi^*(x_1+y_1/2,\cdots,x_n+y_n/2)
\nonumber\\
&~~~~~~~~~~~~~~~~\times\psi(x_1-y_1/2,\cdots,x_n-y_n/2),
\end{align}
in terms of its wave function $\psi(x_1,\cdots,x_n)$.

Following the derivation given in Refs.~\cite{Chen:2003tn,Chen07},
where the non-relativistic limit is taken and the hadron wave
functions are assumed to be those of a spherically symmetric
harmonic oscillator, the above equation can be reduced  to
\begin{eqnarray}\label{ncoal}
N_h^{\rm coal}=g_h \prod_{j=1}^n \frac{N_j}{g_j} \prod_{i=1}^{n-1}
{\int d^3 y_i d^3k_i f_i(k_i) f^W(y_i,k_i) \over \int d^3 y_i d^3
k_i f_i(k_i)}, \label{Eq:Coal}
\end{eqnarray}
where $f^W(y_i,k_i)$ with $y_i$ and $k_i$ being, respectively, the
internal (relative) spatial and momentum coordinates is the Wigner
function associated with the internal (relative) wave function. As
in Ref.~\cite{Chen07}, we assume that the particles in the
emission source are uniformly distributed in space and have
momentum distributions $f_j(p_j)$ given by the Boltzmann
distribution of temperature $T$ only for the transverse momentum
$p_{j,T}$, while the strong Bjorken correlation of equal spatial
($\eta_j$) and momentum ($Y_j$) rapidities is imposed for the
longitudinal momentum, i.e.,
\begin{align}
\label{Eq:therm} f_j(p_j) \propto \delta(Y_j-\eta_j) \exp
\left(-\frac{p_{j,T} ^2}{2m_j T}\right),
\end{align}
with $\eta_j = \log[(t_j+z_j)/(t_j-z_j)]/2$ and $Y_j =
\log[(E_j+p_{j,z})/(E_j-p_{j,z})]/2$ being the space-time and
momentum-energy rapidities, respectively. From the relation
\begin{align}
\prod_{j=1}^n \exp \left(-\frac{p_{j,T} ^2}{2m_j T}\right)
 = \exp\left (-\frac{P_T ^2}{2MT}\right) \prod_{i=1}^{n-1} \tilde{f}_i(k_i),
\end{align}
with $P_T$ and $M$ denoting the total transverse momentum and the
total mass, respectively, we then obtain the 2-dimensional
momentum distribution function of the constituents in Jacobi
coordinates, $\tilde{f}_i(k_i)$,
\begin{eqnarray}
\tilde{f}_i(k_i) \propto \exp\left(-\frac{k_{i}^2}{2\mu_i
T}\right),
\end{eqnarray}
where the reduced constituent masses $\mu_i$ are defined by
\begin{equation}
\frac{1}{\mu _{i}}=\frac{1}{m_{i+1}}+\frac{1}{\sum_{j=1}^{i}{m_{j}}},
\label{reduced-mass1}
\end{equation}
or explicitly
\begin{align}
&\mu_1=\frac{m_1 m_2}{m_1+m_2}, ~~\mu_2=\frac{m_3(m_1+ m_2)}{m_1+m_2+m_3},
\nonumber\\
&\mu_3=\frac{m_4(m_1+m_2+ m_3)}{m_1+m_2+m_3+m_4},~~
\nonumber\\
&\mu_4=\frac{m_5(m_1+m_2+m_3+ m_4)}{m_1+m_2+m_3+m_4+m_5},~~{\rm etc.} \label{reduced-mass2}
\end{align}

In the non-relativistic limit, the rapidity variables are
simplified at midrapidities ($Y = \eta \sim 0$) as $\eta_j \simeq
z_j/t_j$ and $Y_j \simeq p_{j,z}/m_j$. We can thus omit the
contribution from the longitudinal momentum in the Wigner function
$f^W$ as long as the time where the coalescence takes place after
the collision is large compared with the internal time scale of
the hadron, $t_j \gg 1/\omega$, where $\omega$ is the oscillator
frequency. In this case, the 3-dimensional momentum integrations
in Eq.(\ref{ncoal}) reduces to 2-dimensional ones over
$\tilde{f}_i(k_i)$ and the Wigner functions in transverse momentum
$k_i$. The latter are given explicitly as
\begin{align}
\label{Eq:spdWig}
f^W_s(y_i,k_i) = & 8 \exp\left( -\frac{y_i^2}{\sigma_i^2}-k_i^2
\sigma_i^2\right), \nonumber \\
f^W_p(y_i,k_i) = & \bigg( \frac{16}{3} \frac{y_i^2}{\sigma_i^2}
-8+\frac{16}{3} \sigma_i^2 k_i^2 \bigg) \nonumber \\
&\times \exp\left( -\frac{y_i^2}{\sigma_i^2}-k_i^2
\sigma_i^2\right)
\nonumber\\
f^W_d(y_i,k_i) =&
\frac{16}{30}\bigg[4\frac{y_i^4}{\sigma_i^4}-20\frac{y_i^2}{\sigma_i^2}
+15-20\sigma_i^2k_i^2+4\sigma_i^4k_i^4  \nonumber \\
+& 16y_i^2k_i^2-8(\vec y_i\cdot\vec k_i)^2\bigg]\exp{\Big(
-\frac{y_i^2}{\sigma_i^2}-k_i^2\sigma_i^2\Big)},\nonumber\\
\end{align}
for the $s$-wave, $p$-wave, and $d$-wave, respectively, with the
parameters $\sigma_i=1/\sqrt{\mu_i \omega}$ related to the
oscillator frequency $\omega$ and the reduced constituent masses
$\mu_i$.
In Appendix A, we derive the Wigner function for the $d-$wave
state from the harmonic oscillator wave functions.

Carrying out the phase-space integrals in Eq.~(\ref{ncoal}) as
shown in Appendix B, we obtain the coalescence factor for each
relative coordinate,
\begin{align}
F(\sigma_i, \mu_i, l_i, T)
\equiv& \frac{\int d^3y_i d^2k_i \tilde{f}_i(k_i)f^W (y_i,k_i)}
      {\int d^2 k_i \tilde{f}_i(k_i)} \nonumber \\
=&\frac{(4\pi \sigma_i ^2)^{3/2}}{1+2\mu_i T \sigma_i ^2}
\frac{(2 l_i) !!}{(2l_i +1)!!}
\left[\frac{2\mu_i T \sigma_i ^2}{1+2\mu_i T \sigma_i ^2}\right]^{l_i},
\label{cf}
\end{align}
where $l_i$ is the angular momentum of the wave function
associated with the relative coordinate $y_i$. Combining these
results, we obtain the following simple expression for the yield
of hadrons from the coalescence model:
\begin{align}\label{nquark}
N_h^{\rm coal}
\simeq&
gV \prod_{j=1}^n \frac{N_j}{g_j V} \prod_{i=1}^{n-1} F(\sigma_i, \mu_i, l_i, T)
\nonumber\\
\simeq& gV \prod_{j=1}^n \frac{N_j}{g_j V}
\prod_{i=1}^{n-1}
\frac{(4\pi\sigma_i^2)^{3/2}}{1+2\mu_iT\sigma_i^2} \nonumber \\
\times& \frac{(2l_i)!!}{(2l_i+1)!!} \left[ \frac{(2\mu_i
T\sigma_i^2)}{(1+2\mu_{i}T\sigma_{i}^{2})} \right]^{l_i}
\nonumber \\
\simeq& \frac{gV(M\omega)^{3/2}}{(4\pi)^{3/2}}
\frac{(2T/\omega)^L}{(1+2T/\omega)^{n+L-1}} \nonumber \\
\times& \prod_{j=1}^n \frac{N_j(4\pi)^{3/2}}{g_j V
(m_j\omega)^{3/2}}\prod_{i=1}^{n-1} \frac{(2l_i)!!}{(2l_i+1)!!},
\end{align}
where $l_i$ is $0$ for an $s$-wave, $1$ for a $p$-wave and $2$ for
a $d$-wave constituent, $L=\sum_{i=1}^{n-1} l_i$, and
$M=\sum_{i=1}^n m_i$. Here we have used the relation
$\mu_i\sigma_i^2=1/\omega$ to convert the main dependence on $l_i$
into the form of the orbital angular momentum sum $L$. For
$L\geq2$, the last factor in Eq.(\ref{nquark}) depends on the way
$L$ is decomposed into $l_i$. For example, for $L=2$ and $n=3$,
the combination $(l_1, l_2)=(1,1)$ gives a factor $4/9$, while
$(l_1, l_2)=(2,0)$ leads to a factor $8/15$.

\subsubsection{Quark coalescence}

\begin{table}[htdp]
\caption{Yields of normal hadrons at RHIC and LHC in the
coalescence and statistical models with oscillator frequencies
$\omega=550~\MeV, \omega_s=519~\MeV,
\omega_c=385~\MeV$, and $\omega_b=338~\MeV$
are determined by fitting the statistical model results
for $\Lambda(1115)$, $\Lambda_c(2286)$, and $\Lambda_b(5620)$
marked with (*) at RHIC after taking account of resonance decays.
Numbers in the parentheses are those without the decay
contribution.}
\begin{center}
\begingroup
\renewcommand{\arraystretch}{1.2}
\begin{tabular}{c|c|cc|cc}
\hline \hline
\multirow{2}{*}{config.}
& \multirow{2}{*}{particle}
& \multicolumn{2}{|c}{RHIC}
& \multicolumn{2}{|c}{LHC}
\\
\cline{3-6}
& & coal.  & stat.  & coal.  & stat.
\\
\hline
\multirow{4}{*}{$\bar{\rm q}$q}
 & $\omega(782)$    & 44.2  & 40.2  & 119   & 108   \\
 & $\rho(770)$      & 132   & 127   & 358   & 342   \\
 & $\bar{K}^{\ast}(892)$& 41.2  & 47.2  & 111   & 135   \\
 & $K^{\ast}(892)$  & 41.2  & 52.9  & 111   & 135   \\
\hline
\multirow{2}{*}{qqs}
 & $\Lambda(1115)$  &29.8~(*)&  29.8& 80.5  &  77.5 \\
 &          & (3.0) &(6.5)  &(8.1)  & (16.5) \\
 & $\Lambda(1520)$  & 1.6   & 1.9   & 4.4   & 4.8 \\
\hline
\multirow{2}{*}{qqQ}
 & $\Lambda_{c}(2286)$  &0.60~(*)& 0.60 & 4.0   & 3.6 \\
 &          &(0.058)&(0.14) &(0.39) & (0.83) \\
 & $\Lambda_{b}(5620)$
    & $3.6\times10^{-3}$(*) & $3.6\times10^{-3}$    & 0.14  & 0.13  \\
 &  &($3.6\times10^{-4}$)   &($9.2\times10^{-4}$)   &(0.014)& 0.033 \\
\hline \hline
\end{tabular}
\endgroup
\end{center}
\label{normal}
\end{table}%

To apply the coalescence model to hadron production from the QGP
at the critical temperature $T_c$ when the volume is $V_C$, we
need to fix the oscillator frequency appropriately. This is done
by choosing the oscillator frequencies for light, strange,
charmed, and bottom hadrons ($\omega, \omega_s, \omega_c$ and
$\omega_b$) in the quark coalescence to reproduce the yields of
reference normal hadrons in the statistical model. These values
are then used to predict the yields of exotic hadrons.

For hadrons composed of light (up and down) quarks, we take the
oscillator frequency $\omega=550$ MeV to obtain in the coalescence
model similar $\omega$ and $\rho$ yields as in the statistical
model as shown in Table.~\ref{normal}.

For hadrons composed of light and strange quarks, the parameter
$\omega_s$ in the coalescence model is determined by fitting the
statistical model prediction for $\Lambda(1115)$ including the
contribution from resonance decays. Taking into account states in
the octet and decuplet representations that decay dominantly to
$\Lambda(1115)$, we obtain the following result for heavy ion
collisions at RHIC:
\begin{eqnarray}
N_{\Lambda(1115)}^\mathrm{stat,total} &
=& N_{\Lambda(1115)}^\mathrm{stat}+\frac13 N_{\Sigma(1192)}^\mathrm{stat}
+N_{\Xi(1318)}^\mathrm{stat}
\nonumber\\
&&
+\left(0.87+\frac{0.11}{3}\right)N_{\Sigma(1385)}^\mathrm{stat}
\nonumber \\
& &
+N_{\Xi(1530)}^\mathrm{stat}
+N_{\Omega^{-}(1672)}^\mathrm{stat} \nonumber \\
&=&6.46+\frac13\times13.57+4.73 \nonumber \\
&&
+\left(0.87+\frac{0.11}{3}\right)\times10.91
+3.42+0.81 \nonumber \\ &=&  29.8. \label{stat-feed1}
\end{eqnarray}
In the above formula, 0.87 and $0.11/3$ in the bracket represent,
respectively, the branching ratios of $\Sigma(1385)\rightarrow
\Lambda +\pi$ and $\Sigma(1385) \rightarrow \Sigma^0 + \pi $ in
the $\Sigma(1385)$ decay. All numbers are calculated at $T_H$ and
$V_H$ with $\mu_s=10$ MeV and $\mu_B=20$ MeV for RHIC. To
reproduce the total yield within the coalescence model with the
constituent quark masses $m_{u,d}=300$ MeV and $m_s=500$ MeV, we
need $\omega_s=519$ MeV after taking into account the same
feed-down contributions as in Eq.~(\ref{stat-feed1}).
Specifically, we have from the coalescence model
\begin{align}
N_{\Lambda(1115)}^\mathrm{coal,total}
=&3.01+\frac13\times9.03+2.20 \nonumber \\
 &+\left(0.87+\frac{0.11}{3}\right)\times18.07
+4.40+0.78 \nonumber \\
=&  29.8. \label{coal-feed1}
\end{align}
We will use this parameter to estimate the yield of other hadrons
that are composed of light quarks and strange quarks. As shown in
Table~\ref{normal}, this value leads to a yield of
$\Lambda(1520)$, which has the $s$ quark in the $p$-wave state, in
the coalescence model that is smaller than that in the statistical
model as first pointed out in Ref.~\cite{KanadaEn'yo:2006zk}.

The oscillator frequency for charmed hadrons is fixed to reproduce
the $\Lambda_c(2286)$ yield including the feed-down
contribution~\cite{Oh:2009zj} but without taking into
consideration the effect of the diquarks~\cite{Lee:2007wr}. For
the $\Lambda_c(2286)$ yield, we consider only the contribution
from $\Sigma_c(2455)$, $\Sigma_c(2520)$ and $\Lambda_c(2625)$
decays as states of higher masses are negligible, that is,
\begin{eqnarray}
N_{\Lambda_c(2286)}^\mathrm{stat,total}
&=&  N_{\Lambda_c(2286)}^\mathrm{stat}
     +N_{\Sigma_c(2455)}^\mathrm{stat}
    +N_{\Sigma_c(2520)}^\mathrm{stat} \nonumber \\ &&
    +0.67\times N_{\Lambda_c(2625)}^\mathrm{stat} \nonumber \\
&=& 0.139+0.177+0.254+0.67 \times0.048 \nonumber \\
&= & 0.602 \label{stat-feed2}
\end{eqnarray}
at RHIC. Fitting again the total yield of $\Lambda_c(2286)$
calculated in the statistical model to that in the  coalescence
model for the same resonances as given in Eq.~(\ref{stat-feed2}),
\begin{align}
N_{\Lambda_c(2286)}^\mathrm{coal,total}
=& 0.058+0.173+0.346+0.67 \times0.037 \nonumber \\
=& 0.602, \label{coal-feed2}
\end{align}
we obtain $\omega_c=385$ MeV for the charm quark mass $m_{c}=1500$
MeV.

The oscillator frequency for bottom hadrons $\omega_b=338~\MeV$ is
obtained by fitting the sum of the statistical model results for
$\Lambda_b(5620)$ and the contribution from $\Sigma_b(5810)$ and
$\Sigma_b^*(5830)$ decays at RHIC,
\begin{align}
N_{\Lambda_b(5620)}^\mathrm{stat,total}
=&
  N_{\Lambda_b(5620)}^\mathrm{stat}
 +N_{\Sigma_b(5810)}^\mathrm{stat}
 +N_{\Sigma_b(5830)}^\mathrm{stat}
\nonumber\\
=& \Exm{9.2}{-4}+\Ex{9.7}{-4}+\Ex{1.73}{-3}
\nonumber\\
= & \Exm{3.62}{-3}
\ , \label{stat-feed3}
\\
N_{\Lambda_b(5620)}^\mathrm{coal,total}
=& \Exm{3.62}{-4}+\Ex{1.085}{-3}+\Ex{2.170}{-3}
\nonumber\\
= & \Exm{3.62}{-3}
\ . \label{coal-feed3}
\end{align}
with the bottom quark mass $m_{b}=4700$ MeV.

Since the oscillator frequencies are related to the sizes of
hadrons \cite{Chen:2003tn,Chen07}, same values as determined at
RHIC are used in the coalescence calculations for heavy ion
collisions at LHC. Using the same $\omega$ values as those for
normal hadrons, we then see from Eq.~(\ref{nquark}) that the
addition of a $s$-wave, $p$-wave, or $d$-wave quark leads to,
respectively, a coalescence factor
\begin{align}
\frac{1}{g_i}\frac{N_i}{V} \frac{(4 \pi \sigma_i^2)^{3/2} }{(1+2
\mu_i T \sigma_i^2)}  \sim & 0.360
\nonumber \\
\frac{1}{g_i}\frac{N_i}{V} \frac{2}{3} \frac{(4 \pi
\sigma_i^2)^{3/2} 2 \mu_i T\sigma_i^2 }{(1+2 \mu_i T
\sigma_i^2)^2} \sim & 0.093
\nonumber \\
\frac{1}{g_i}\frac{N_i}{V} \frac{8}{15} \frac{(4 \pi
\sigma_i^2)^{3/2} (2 \mu_i T\sigma_i^2)^2 }{(1+2 \mu_i T
\sigma_i^2)^3} \sim & 0.029. \label{coal-factors}
\end{align}
The production of multiquark hadrons involves more $s$-, $p$- and
$d$-wave coalescence factors and is hence generally suppressed.
Moreover, the $d$-wave coalescence is suppressed in comparison
with the $p$-wave coalescence, which is further suppressed
relative to the $s-$wave coalescence~\cite{KanadaEn'yo:2006zk}.

\subsubsection{Hadron coalescence}

For the yields of weakly bound hadronic molecules from the
coalescence of hadrons, they are evaluated at the kinetic
freeze-out temperature $T_{F}$ and volume $V_F$. The oscillator
frequencies needed for hadronic molecules in the hadron
coalescence is related to their mean square distances $\langle
r^2\rangle$ between the two constituent hadrons. For a hadronic
molecule in the relative $s$-wave state, the oscillator frequency
is given by
\begin{align}
\omega = \frac{3}{2\mu_R \langle{r^2}\rangle}
\ ,
\label{Eq:rmsr-omega}
\end{align}
where $\mu_R=m_1 m_2/ (m_1+m_2)$ is the reduce mass. The mean
square distance of the hadronic molecule can be further related to
its binding energy $B$ via the scattering length $a_0$ of the two
interacting constituent hadrons, i.e.,
\begin{align} B
\simeq & \frac{\hbar^2}{2\mu_R a_0^2} \ ,\quad \langle{r^2}\rangle
\simeq \frac{a_0^2}{2} \ , \label{binding-radius}
\end{align}
These relations are valid when the binding energy is small and the
scattering length is large compared to the range of the hadronic
interaction, and they can be easily obtained as follows. Using the
relation $k\cot{\delta_0 (k)} = -1/a_{0}$ for the relative
momentum $k \rightarrow 0$ between the $s$-wave scattering phase
shift $\delta_0$ and the scattering length $a_0$, the S matrix for
two interacting hadrons at low energies can be approximated as
\begin{equation}
S = e^{2i\delta_0 (k)} \approx \frac{-\frac{1}{a_0} + ik}{-\frac{1}{a_0} - ik},
\label{phase-a}
\end{equation}
which has a pole at $k = i/a_0$, corresponding to a bound state
with the binding energy given by the first equation in
Eq.~(\ref{binding-radius}) and the radial wave function outside
the interaction range  $u_b (r) \sim e^{-r/a_{0}}$. Assuming that
$a_0$ is much larger than the interaction range and using the
above form of the wave function for the whole region, we obtain
the mean square distance given by the second equation in
Eq.~(\ref{binding-radius}). For a weakly bound two-body states, we
thus obtain from Eqs.~(\ref{Eq:rmsr-omega}) and
(\ref{binding-radius}) the simple relation $\omega=6B$. We note
that $\langle{r^2}\rangle$ is the mean square distance in the
relative coordinate, and it is not the squared mean radius from
the center-of-mass.

For example, the oscillator frequency for $f_0(980)$ can be
obtained from $\omega_{f_0(980)}=6\times B_{f_0(980)}=67.8$ MeV
with $B_{f_0(980)}=M_{K^+}+M_{\bar{K}_0}-M_{f_0(980)}
=493.7+497.6-980=11.3$ MeV. As another example, the oscillator
frequency $\omega_{\Omega\Omega}$ for the diomega $(\Omega
\Omega)_{0^{+}}$ predicted by the chiral quark
model~\cite{di-omega} can be calculated from
Eq.~(\ref{Eq:rmsr-omega}),
\begin{eqnarray}
\omega_{\Omega\Omega} &=& \frac{3}{2\mu_{\Omega\Omega}
\langle{r^2}\rangle_{\Omega\Omega}}
\nonumber \\
&=&\frac{3}{2}\frac{197.3^2}{1672.45/2 \times 0.84^2}
=98.8~\MeV
\end{eqnarray}
where
$\sqrt{\langle{r^2}\rangle}_{\Omega\Omega}=0.84~\mathrm{fm}$~\cite{di-omega}
has been used. By the same token, we can calculate the oscillator
frequencies for all the hadronic molecules, and the results are
summarized in Table \ref{summary}.

\begin{table*}[htdp]
\caption{List of exotic hadrons discussed in this paper. Shown are
the mass ($m$), degeneracy ($g)$, isospin ($I$), spin and parity
($J^P$), the quark structure ($2q/3/q/6q$ and $4q/5q/8q$),
molecular configuration (Mol.) and corresponding oscillator
frequency ($\omega_{\rm Mol.}$), and decay mode of a hadron. For
the $\omega_\mathrm{Mol.}$, it is fixed by the binding energies B
of hadrons ($\omega \simeq 6 \times \mathrm{B}$, marked (B)) or
their mean square distances $\langle r^2\rangle$ ($\omega \simeq
3/2\mu\VEV{r^2}$, marked (R)). In the case of three-body molecular
configurations for exotic dibaryons, we adopt the
$\omega_\mathrm{Mol.}$ as that for the subsystem, as marked (T).
Further marked by $^{*)}$ are undetermined quantum numbers of
existing particles, by $^{\ddag)}$ particles which are not yet
established, and by $^{\dag)}$ particles which are newly predicted
by theoretical models.}
\begin{tabular}{c|cccc|c|c|c|c|c}
\hline
\hline
  \parbox[c]{1.5cm}{Particle}
& \parbox[c]{1.0cm}{$m$ (MeV)} & \parbox[c]{0.3cm}{$g$} &
\parbox[c]{0.5cm}{$I$} & \parbox[c]{0.8cm}{$J^P$} &
\parbox[c]{1.3cm}{$2q/3q/6q$}
& \parbox[c]{1.5cm}{~\\[-0.5ex]$4q/5q/8q$\\[0.5ex]}
& \parbox[c]{1.1cm}{Mol.}
& \parbox[c]{1.5cm}{~\\[-0.3ex]$\omega_\mathrm{Mol.}$ (MeV)\\[0.5ex]}
& \parbox[c]{1.5cm}{decay mode}
\\
\hline
Mesons &&&&&&&&&\\

$f_0(980)$  & 980 & 1 & 0 & $0^+$ & $q\qbar, s\sbar~(L=1)$ &
$q\qbar s\sbar$ & $\bar{K}K$  & 67.8(B)  & $\pi\pi$ (strong decay) \\

$a_0(980)$  & 980 & 3 & 1 & $0^+$ & $q\qbar~(L=1)$  & $q\qbar
s\sbar$ & $\bar{K}K$ & 67.8(B) & $\eta\pi$ (strong decay) \\

$K(1460)$ & 1460 & 2 & 1/2 & $0^-$ &
$q\bar s$ & $q \bar q q \bar s$ & $\bar KKK$ & 69.0(R) & $K\pi\pi$ (strong decay) \\

$D_s(2317)$ &2317& 1&0  &$0^+$   & $c\sbar~(L=1)$  & $q\qbar
c\sbar$ & $DK$ & 273(B)   & $D_{s}\pi$ (strong decay) \\

$T_{cc}^1$ $^{\dag)}$  &3797& 3&0  &$1^+$ & --- & $qq\cbar\cbar$
& $\bar{D}\bar{D}^*$    & 476(B)&$K^{+}\pi^{-}+K^{+} \pi^{-}+\pi^{-}$ \\

$X(3872)$ & 3872 & 3 & 0 & $1^+$, $2^-$ $^{*)}$ & $c\cbar~(L=2)$ &
$q\qbar c\cbar$ & $\bar{D}D^*$
& 3.6(B)  & $J/\psi\pi\pi$ (strong decay) \\

$Z^+(4430)$ $^{\ddag)}$ &4430& 3&1 &$0^-$ $^{*)}$ & --- & $q\qbar
c\cbar~(L=1)$ & $D_1\bar{D}^*$ & 13.5(B) & $J/\psi\pi$ (strong decay) \\

$T_{cb}^0$ $^{\dag)}$ & 7123 & 1 & 0 & $0^+$ & --- &
$qq\cbar\bbar$
& $\bar{D}B$ & 128(B) & $K^{+}\pi^{-}+K^{+}\pi^{-}$ \\

\hline
Baryons &&&&&&&&&\\

$\Lambda(1405)$ & 1405 & 2 & 0 & $1/2^-$ & $qqs~(L=1)$ &
$qqqs\qbar$ &
$\bar{K}N$ & 20.5(R)-174(B) & $\pi\Sigma$ (strong decay) \\

$\Theta^+(1530)$ $^{\ddag)}$ & 1530 & 2 & 0 & $1/2^+$ $^{*)}$ &
--- &
$qqqq\sbar~(L=1)$ & --- & --- & $KN$ (strong decay) \\

$\bar{K}KN$ $^{\dag)}$ & 1920 & 4 & 1/2 & $1/2^+$ & --- &
$qqqs\sbar~(L=1)$ & $\bar{K}KN$ & 42(R) & $K\pi\Sigma$, $\pi \eta N$ (strong decay) \\

$\bar{D}N$ $^{\dag)}$  & 2790 & 2 & 0 & $1/2^-$ & --- &
$qqqq\cbar$ &
$\bar{D}N$ & 6.48(R) & $K^{+}\pi^{-}\pi^{-}+p$ \\

$\bar{D}^*N$ $^{\dag)}$  & 2919 & 4 & 0 & $3/2^-$ & --- &
$qqqq\cbar~(L=2)$ &
$\bar{D}^{\ast}N$ & 6.48(R) & $\bar{D}+N$ (strong decay) \\

$\Theta_{cs}$ $^{\dag)}$ & 2980 & 4 & 1/2 & $1/2^+$ & --- &
$qqqs\cbar~(L=1)$ & --- & --- & $\Lambda+K^{+}\pi^{-}$ \\

$BN$ $^{\dag)}$ & 6200 & 2 & 0 & $1/2^-$ & --- & $qqqq\bbar$ &
$BN$ &
25.4(R) & $K^{+}\pi^{-}\pi^{-}+\pi^{+}+p$ \\

$B^*N$ $^{\dag)}$  & 6226 & 4 & 0 & $3/2^-$ & --- &
$qqqq\bbar~(L=2)$ &
$B^{\ast}N$ & 25.4(R) & $B+N$ (strong decay) \\

\hline
Dibaryons &&&&&&&&&\\

$H$  $^{\dag)}$ & 2245 & 1 & 0 & $0^+$ & $qqqqss$ & --- & $\Xi{N}$
&
73.2(B) & $\Lambda \Lambda$ (strong decay) \\

$\bar{K}NN$ $^{\ddag)}$ & 2352 & 2 & 1/2 & $0^-$ $^{*)}$ &
$qqqqqs~(L=1)$ & $qqqqqq\,s\qbar$ & $\bar{K}NN$ & 20.5(T)-174(T)
& $\Lambda N$ (strong decay) \\

$\Omega\Omega$ $^{\dag)}$ & 3228 & 1 & 0 & $0^+$ & $ssssss$ & ---
&
$\Omega\Omega$ & 98.8(R) & $\Lambda K^{-}+\Lambda K^{-}$ \\

$H_c^{++}$ $^{\dag)}$ & 3377 & 3 & 1 & $0^+$ & $qqqqsc$ & --- &
$\Xi_cN$ & 187(B) & $\Lambda K^{-}\pi^{+}\pi^{+}+p$ \\

$\bar{D}NN$ $^{\dag)}$ & 3734 & 2 & 1/2 & $0^-$ & --- &
$qqqqqq\,q\cbar$ & $\bar{D}NN$ & 6.48(T) & $K^{+}\pi^{-}+d$,
$K^{+}\pi^{-}\pi^{-}+p+p$ \\

$BNN$ $^{\dag)}$ & 7147 & 2 & 1/2 & $0^-$ & --- & $qqqqqq\,q\bbar$
&
$BNN$ & 25.4(T) & $K^{+}\pi^{-}+d$, $K^{+}\pi^{-}+p+p$ \\

\hline
\hline
\end{tabular}
\label{summary}
\end{table*}

\section{Exotic hadrons}\label{properties}

In this Section, we briefly discuss the properties, such as the quantum numbers
and possible decay modes, and the current theoretical and experimental status
of the exotic hadrons included in the present study. For convenience, we
classify these hadrons into exotic mesons (Subsec. A), exotic
baryons (Subsec. B) and exotic dibaryons (Subsec. C) as shown
in Table~\ref{summary}.

\subsection{Exotic mesons}

For exotic mesons, we include the following:

\begin{enumerate}

\item $f_0(980)$: This $I=0$ scalar particle together with the
$I=1$ $a_0(980)$ are members of the scalar nonet that has been
thought to be composed of multiquark configurations
\cite{Jaffe76-1,Jaffe76-2}. If $f_0(980)$ is a member of the
multiquark configurations, its wave function then has a hidden
$\bar{s}s$ component. Assuming that they are composed of
quark-antiquark pair, their wave functions would be $f_0(980) \sim
s \bar{s}$ and $a_0(980) \sim (u \bar{u} -d \bar{d})/\sqrt{2}$. An
early QCD sum rule analysis suggested, on the other hand, that
$f_0(980)$ and $a_0(980)$ were just the $I=0$ and $1$ combinations
of $u \bar{u}$ and $d \bar{d}$ \cite{RRY84}. There are also models
in which $f_0(980)$ is a $K \bar{K}$ molecule. While there seems
to be consensus from lattice calculations that this particle is a
tetraquark state, the situation is not at all clear because of the
treatment of disconnected diagrams \cite{Alford00}.

\item $K(1460)$: This is an exited state of kaon with $J^P = 0^-$
on the particle list of the Particle Data Group (PDG) but is
omitted from the summary table~\cite{Nakamura:2010zzi}. It was
observed in the partial wave analysis of the $K \pi \pi$ final
state in elementary reactions. Recently, this resonance has been
studied theoretically in the context of the meson dynamics. In
Ref.~\cite{Albaladejo:2010tj} this kaon was obtained from the
$K$-$f_0(980)$ $s$-wave two-body dynamics with the $f_0(980)$
dynamically generated in the $\bar KK$ and $\pi\pi$ coupled
channels. A later study in Ref.~\cite{Torres:2011jt} considered
the three-body coupled-channel dynamics of $\bar KKK$ in a Faddeev
approach and found a very similar kaonic excitation. A
non-relativistic potential model for the $\bar KKK$ system was
also used in Ref.~\cite{Torres:2011jt}, and a quasi-bound state of
binding energy 21 MeV and root mean square radius 1.6 fm was
obtained. Thus, the $K(1460)$ could be understood as a $\bar KKK$
hadronic molecular state.

\item $D_{sJ}(2317)$:  This state was first observed by the BaBar
Collaboration \cite{Aubert:2003fg} through the $D_s^+ \pi^0$
channel in inclusive $e^+e^-$ annihilation. Its measured mass is
approximately 160 MeV below the prediction of the very successful
quark model for the charmed meson \cite{Godfrey}. Due to its low
mass, the structure of $D_{sJ}(2317)$ has been under extensive
debate. It has been interpreted as a $c \bar{s}$ state
\cite{Dai,Bali,Dougall,Hayashigaki,Narison}, two-meson molecular
state \cite{Barnes,Szczepaniak}, $D-K$ mixing \cite{Beveren},
four-quark state \cite{Cheng,Terasaki,Maiani,Bracco} or a mixture
of two-meson and four-quark states \cite{Browder}.

\item $T^1_{cc}(ud\bar{c}\bar{c})$: The set of tetraquarks with
two heavy quarks were first considered in Ref.~\cite{Zouzou86}.
The structure with [ud] diquark is expected to be particularly
stable~\cite{Wilczek,Tsushima} and could be bound against the
strong decay into $D_1 D$. The quantum number is $J^P=1^+$ with
$I=0$; hence the decay into $DD$ is forbidden due to the angular
momentum conservation. Estimates based on the simple color-spin
interaction suggest the mass to be 3796 MeV~\cite{Lee07,Lee09}.
The hadronic decay mode of $T_{cc}$ is $D^{*-}\bar{D}^0$ if its
mass is above the threshold and $\bar{D}^0 D^0 \pi^-$ if below. If
the $T_{cc}$ is strongly bound, it can then decay weakly to
$D^{*-} K^+ \pi^-$ with a lifetime similar to that of the
$\bar{D}$ meson. A molecular state with the same quantum number
was also predicted to exist within the pion-exchange model
\cite{Manohar:1992nd}. The production of doubly charmed hadrons in
heavy ion collisions at RHIC was discussed in
Ref.~\cite{Schaffner}, and also estimated for LHC~\cite{Lee07}.
This state could also be searched for at Belle. In particular, the
search should be similar to that for the doubly charmed baryon
such as the $\Xi_{cc}$. Here are the two possibilities. The first
one is through the decay of the $B$ meson. Unfortunately the
search at Belle was not successful~\cite{Belle-cc} as the dominant
subprocess was the weak decay of  the $\bar{b}$ quark into a
$\bar{c}$ by emitting a $W^+ (\rightarrow c \bar{s})$. Therefore,
while the $c \bar{c}$ is produced, the $cc$ creation might be
highly suppressed. On the other hand, another more feasible search
is in the continuum background where two $c\bar{c}$ pairs are
known to be produced in the reaction $e^+e^- \rightarrow J/\psi
X(3940)$ \cite{Belle-X3940}.

\item $X(3872)$:  The Belle collaboration found this particle in
the $B^+\rightarrow X(3872) K^+ \rightarrow J/\psi \pi^+ \pi^-
K^+$ decay \cite{Choi:2003ue}.  CDF, D0, and BaBar have confirmed
its existence, and the current world average mass is $3871.2 \pm
0.39$ MeV. Although the new BaBar result favors the
$J^{PC}=2^{-+}$ assignment \cite{newbabar}, the established
properties of the $X(3872)$ are in conflict with this assignment
\cite{kane,bpps}. Therefore, the favored quantum numbers are
$J^{PC}=1^{++}$ with the isospin violating decay modes. This
particle was predicted in Ref.~\cite{Tornqvist} as a $D
\bar{D}^{*0}$ bound state within a meson-exchange model. It was
shown in this study that in the $D \bar{D}^{*}$ sector the
one-pion exchange interaction alone is strong enough to form a
molecular state that is bound by approximately 50 MeV. In this
case, other molecular states of $D^*\bar{D}^{*}$, $D_1\bar{D}^*$,
$D_1 \bar{D}$ and $D_0\bar{D}$ are also expected to exist as the
pion exchange is allowed in these channels as well. In other
studies, it has been claimed that the $X(3872)$ has the admixture
of the $\bar{c}c$ component and is thus a tetraquark
hadron~\cite{Marina-review}. Its dominant decay modes include
$J/\psi \pi^+ \pi^-$, $J/\psi \pi^+ \pi^-$, and $D^0 \bar{D}
\pi^0$.

\item $Z^+(4430)$: The Belle collaboration observed this charged
state in $B^+ \rightarrow K \psi' \pi^+$ through its decay into
$\psi\prime\pi^+$ \cite{Belle:Z4430}. The reported mass and width
are $M=4433$ MeV and $\Gamma =45 ^{+18+30}_{-13-13}$ MeV,
respectively.  The reported mass is close to the $D_1\bar{D}^*$
threshold and hence the possible structure for this state is
either a molecular or a tetraquark state \cite{Marina-review}.  As
commented in the discussion on $X(3872)$, the one-pion exchange
interaction could bound a $D_1\bar{D}^*$ molecular state. So far,
no other experiment has confirmed this finding. In particular,
BaBar \cite{babarz} also searched the $Z^-(4430)$ signature in
four decay modes: $B\to\psi\pi^-K$, where $\psi=J/\psi$ or
$\psi^\prime$ and $K=K_S^0$ or $K^+$. No significant evidence for
a signal peak was found in any of the investigated processes.
After the failure of the BaBar collaboration in confirming the
$Z^-(4430)$ mass peak, Belle has performed a reanalysis of their
data that took detailed account of possible reflections from the
$K\pi^-$ channel. From a full Dalitz plot reanalysis of their
data, Belle has confirmed the observation of the $Z^+(4430)$
signal with a 6.4$\sigma$ peak significance. The updated
$Z^+(4430)$ parameters are: $M=(4433^{+15+19}_{-12-13})\MeV$ and
$\Gamma=(109^{+86+74}_{-43-56})\MeV$ \cite{bellezn}. If confirmed,
the $Z^+(4430)$ is the first prime candidate for an exotic
particle. Considering the $Z^+(4430)$ as a loosely bound $s$-wave
$D_1\bar{D}^*$ molecular state, the allowed angular momentum and
parity are $J^{P}=0^-,1^-,2^-$, although the $2^-$ assignment is
probably suppressed in the $B^+ \rightarrow Z^+ K$ decay by the
small phase space. Among the remaining possible $0^-$ and $1^-$
states, the former will be more stable as the latter can also
decay to $D_{1} \bar{D}$ in $s$-wave. Hence the $0^-$ quantum
number is favored.

Very recently the BELLE collaboration \cite{belle_zb} reported the
observation of two narrow charged structures in the hidden-bottom
decay channels $\pi^{\pm}\Upsilon(nS)~(n=1,2,3)$ and
$\pi^{\pm}h_b(mP)~(m=1,2)$ of $\Upsilon(5S)$. The measured masses
(widths) of these two structures are, in units of MeV,
$M_{Z_{b}}=10610$ ($\Gamma_{Z_{b}}= 15.6 \pm 2.5$) and
$M_{Z'_{b}}=10650$ ($\Gamma_{Z'_{b}}= 14.4 \pm 3.2$),
respectively. The analysis of the $Z_b$ states decay in the
channel $Z_b^\pm \rightarrow \Upsilon(2S) \pi^\pm$ favors the
$I^G(J^P) = 1^+(1^+)$ assignment. Since the masses of these two
states are very  close to the $\bar{B}B^*(10604.6\mbox{ MeV})$ and
$\bar{B}^*B^*(10650.2\mbox{ MeV})$ thresholds, they are ideal
candidates for these molecular states. It is interesting to notice
that the decay channels of the $Z_b^\pm$ states are very similar
to the decay channel of $Z^+(4430)$.

\item $T^0_{cb}(ud \bar{c}\bar{b})$: Both the light diquark and
heavy anti-diquark are scalar diquarks so that $J^P=0^+$ with
$I=0$. This particle could be strongly bound with a mass of 7149
MeV\cite{Lee09}.  As it has the $\bar{D}^0 B^0$ component, it can
decay weakly via $T_{cb}^0 \rightarrow K^+ \pi^-+ K^+ \pi^-$.

\end{enumerate}

\subsection{Exotic baryons}

For exotic baryons, we consider the following:

\begin{enumerate}

\item $\Lambda(1405)$: This resonant state with $I=0$,
$J^{P}=1/2^{-}$, mass $1406 \pm 4$ MeV, and width $50 \pm 2$
MeV~\cite{Nakamura:2010zzi} has been considered as a quasi-bound
state of the $\bar{K} N$ system~\cite{Dalitz:1960du}, even before
the establishment of the QCD. The modern theoretical approach
based on the chiral dynamics within the unitary framework (the
chiral unitary
approach)~\cite{Kaiser,Hyodo:2011ur,Oset,Oller:2000fj,Lutz:2001yb,Jido:2003cb,Hyodo:2007jq}
also suggests that this resonant state is dynamically generated in
the meson-baryon scattering including the $\bar{K}N$ and $\pi
\Sigma$ channels, and is dominated by the meson-baryon molecular
component~\cite{Hyodo08}. The mean square distance between
$\bar{K}$ and $N$ in the $\Lambda (1405)$ is evaluated to be
$\langle r^{2}\rangle = 2.7$ fm$^{2}$ in the chiral unitary
approach~\cite{Sekihara}, in which the $\Lambda (1405)$ peak
appears around 1420 MeV in the $\bar{K} N$ channel. This value
leads to a smaller oscillator frequency for the bound state
($\omega _{\mathrm{Mol.}} = 20.5$ MeV) than that fixed by the
binding energy with the $\Lambda (1405)$ mass of 1405 MeV. Its
dominant decay mode is $\pi \Sigma$ in the $I=0$ channel, but
there may be the possibility of observing it in the $\gamma
\Lambda$ decay mode.

\item $\Theta^+(uudd\bar{s})$: This flavor exotic baryon with
strangeness $S=+1$, $J^P=1/2^+$ and $I=0$ was predicted in the
chiral soliton model~\cite{Diakonov97}. The intriguing features
are its light mass of 1540 MeV and narrow width, which partly
motivated the first experimental observation by
LEPS~\cite{Nakano:2003qx}. Although the LEPS result was confirmed
by several other experiments, it was subsequently followed by
negative results with high statistics such as in the high energy
collision experiment by PHENIX~\cite{Pinkenburg04} and the low
energy photoproduction experiment by CLAS~\cite{DeVita06}. On the
other hand, the recent LEPS result maintains a positive
signal~\cite{Nakano08}. The most suitable hadronic decay mode for
its identification in an inclusive experiment is $\Theta^+\to
K^0p$. The production rate in heavy ion collisions has been
estimated in the statistical
model~\cite{Randrup,Becattini,Letessier} and in the coalescence
model~\cite{Chen:2003tn}. In the quark model, the spin-parity of
$\Theta^+$ is $1/2^-$ if all five quarks are in the $s$-wave orbit
but is $1/2^+$ after including the strong diquark
correlation~\cite{Jaffe:2003sg}. The possibility of a $3/2^-$
state has also been proposed to explain its narrow
width~\cite{Hosaka:2004bn}. A recent comprehensive QCD sum rule
study favors, however, the $3/2^+$
assignment~\cite{Gubler:2009iq}.

\item ${\bar K}KN$: The quasibound state of $\bar{K}KN$ was
predicted to be a hadronic molecular state of $N^{*}$ with a mass
of 1910 MeV, $J^P=1/2^+$ and  $I=1/2$ in the variational
calculation using a hadronic two-body potential~\cite{Jido-Enyo}.
This state was confirmed by a coupled-channels Faddeev
calculation~\cite{Martinez}. It has been interpreted as a
coexistence state of the $\Lambda(1405)K$ and $a_{0}(980)N$
clusters, and its main decay modes are thus the $K\pi\Sigma$ from
the $\Lambda(1405)$ decay and the $ \pi \eta N$ from the
$a_{0}(980)$ decay. Since the $\bar{K} KN$ is a hadronic molecular
state, it has a large spatial distribution. The root mean square
radius is found to be 1.7 fm and the interhadron distances are
larger than 2 fm~\cite{Jido-Enyo}.

\item $\bar{D}N (\bar{D}^*N)$ and $BN (B^*N)$: The quark contents
of these hadron molecular states are similar to the
$\Theta_{c(b)}$ but with the different quantum numbers of
$J^P=1/2^-$ and $I=0$. Recently, based on the heavy quark
symmetry, a pion induced bound $\bar{D}N$-$\bar{D^*}N$ molecular
state was predicted to exist with a binding energy of a few MeV
below the threshold of 2806 MeV~\cite{Yasui:2009bz}. This is a
shallow bound state compared to the deeply bound $\Theta_c$ of
about 100 MeV below the threshold. An easily identifiable decay
mode is $K^+\pi^-\pi^- p$. The $BN$ molecule would be more stable,
because the heavy quark symmetry amplifies the strong mixing
between $BN$ and $B^*N$ and thus suppresses the kinetic energy.
The mass of the $BN$ molecule was predicted to be a few tens MeV
below the threshold of 6218 MeV, and the possible weak decay mode
is $K^{+}\pi^{-}\pi^{-}+\pi^{+}+p$. In a more recent study
~\cite{Yamaguchi:2011xb}, the $\bar{D}N$ ($BN$) was also predicted
to have a resonance state between the $\bar{D}N$ ($BN$) and
$\bar{D}^{\ast}N$ ($B^{\ast}N$) thresholds, with the mass 2929
(6226) MeV and quantum numbers $J^{P}=3/2^{-}$ and $I=0$. Similar
to the $\bar{D}N$ ($BN$) bound states, this resonance state is
induced by the pion exchange within the heavy quark symmetry. This
resonance state can also be regarded as a bound state of
$\bar{D}^{\ast}$ ($B^{\ast}$) and $N$ with respect to the
$\bar{D}^{\ast}N$ ($B^{\ast}N$) threshold like that of the
Feshbach resonance. Its decay width of 19 (0.12) MeV to the
$\bar{D}N$ ($BN$) via the strong interaction is very narrow as a
result of the suppression by the $d$-wave centrifugal barrier in
the final $\bar{D}N$ ($BN$) state. We have fixed the oscillator
frequency $\omega$ from the root mean square distance,
$\sqrt{\langle{r^2}\rangle}=3.8 (1.7)~\mathrm{fm}$ for the
$\bar{D}N$ ($BN$) molecular state~\cite{Yasui:2009bz}, and the
same $\omega$ is used for its excited state, $\bar{D}^*N$
($B^*N$).

\item $\Theta_c(uudd\bar{c})$ and $\Theta_b(uudd\bar{b})$: The
bound Skyrmion approach predicted the bound exotic hadron
$\Theta_c$ with the mass of 2650 MeV and quantum numbers
$J^P=1/2^+$ and $ I=0$~\cite{Riska93}. There was one experiment
reporting a positive signal~\cite{H1} at a mass around 3.1 GeV,
but no confirmation exists so far~\cite{Focus}. An easily
identifiable decay mode is $K^+\pi^-\pi^- p$ if the state is
strongly bound and $D^{*-}p$ if it is a resonant state. Similarly,
the $\Theta_b$ mass was predicted to be 5207 MeV with the same
quantum numbers $J^P=1/2^+$ and $I=0$. The possible weak decay
mode is $K^{+}\pi^{-}\pi^{-}+\pi^{+}+p$.

\item $\Theta_{cs}(udus\bar{c})$: In the quark model including the
color-spin interaction, the $J^P=1/2^-$ and $I=0$ five-quark state
can be bound and becomes stable against the strong
decay~\cite{Lipkin,Gignoux}. The mass is predicted to be 2920-2930
MeV, depending on the model parameters. This state was searched
for in the Fermilab E791 experiment through the $\phi\pi p$
mode~\cite{Aitala97} and the $K^{*0}K^- p$ mode~\cite{Aitala99},
and the results are so far negative.

\end{enumerate}

\subsection{Exotic dibaryons}

The exotic dibaryons included in the present study are:

\begin{enumerate}

\item  $H$ dibaryon: This particle was first predicted in
Ref.~\cite{Jaffe76} as a deeply bound state below the
$\Lambda\Lambda$ threshold. Despite extensive searches such as in
the BNL-E885 experiments~\cite{E885}, deeply bound $H$ dibaryons
have not been observed. The discovery of the double-$\Lambda$
hypernucleus ($^6_{\Lambda\Lambda}\mathrm{He}$) in the Nagara
event~\cite{Nagara} finally excluded the possibility of deeply
bound $H$ dibaryon, as the two $\Lambda$ particles can decay
strongly into the core nucleus and the $H$ dibaryon if the $H$
mass is below the $\Lambda\Lambda$ threshold by more than the
$\Lambda\Lambda$ separation energy ($B_{\Lambda\Lambda}=7.25 \pm
0.19^{+0.18}_{-0.11}$ MeV)~\cite{Nagara}. As a result, we now have
only a narrow window for the $H$ particle to be bound, $0 < B_H
\lesssim 7~\mathrm{MeV}$, where $B_H$ is the binding energy of $H$
from the $\Lambda\Lambda$ threshold. The reason why the $H$
particle does not strongly bound may be due to the
instanton-induced determinant (Kobayashi-Maskawa-'t
Hooft)~\cite{KMT} interaction, which acts repulsively in the $H$
channel and may cancel the strong color-spin attraction, as
demonstrated in the quark-cluster model~\cite{OkaTakeuchi}. There
is, however, still a possibility that the $H$ particle exists as a
resonance. In the KEK-E522 experiment, an enhancement in the
$\Lambda\Lambda$ invariant mass spectrum is observed at 10-20 MeV
above the threshold~\cite{E522}, while the significance as a peak
is only around 2$\sigma$. This peak-like enhancement cannot be
explained by the final-state interaction~\cite{OHNSA}. Recent
lattice calculations have suggested that a bound state pole exists
around the SU(3) limit~\cite{H-LQCD}. With the realistic SU(3)
breaking, this bound state pole would be shifted to a weakly bound
state or a resonant state between the $\Lambda\Lambda$ and $\Xi N$
thresholds. If we apply the low energy scattering formula (Eq.
(\ref{binding-radius})), the rms radius of the $H$ resonance as a
bound state of $\Xi N$~\cite{E522} may be evaluated to be
$0.9-1.3$ fm.

\item $\bar{K}NN$: Motivated by the existence of the
$\Lambda(1405)$ resonance below the $\bar{K}N$ threshold, the
possibility of bound $\bar{K}$-nuclear systems was proposed in
Ref.~\cite{Akaishi} based on a phenomenological $\bar{K}N$
potential. Since then the simplest $\bar{K}NN$ system has been
intensively studied both theoretically and experimentally. While
the experiment by FINUDA~\cite{FINUDA} indicates a peak structure
at 2255 MeV in the $\Lambda N$ invariant mass spectrum, the
interpretation of the peak as the $\bar{K}NN$ state is still
controversial~\cite{Magas}. Recent rigorous few-body calculations
for the $\bar{K}NN$ system indicate that the system bounds in the
$J^P=0^-$ and $I=1/2$ channel~\cite{Shevchenko,Ikeda1,Ikeda2,Yamazaki-Akaishi}.
With a suitable treatment of the energy dependence of the $\bar{K}N$
interaction, the mass of the $\bar{K}NN$ system is predicted to be
about 2350 MeV~\cite{Dote1,Dote2,Ikeda:2010tk}. In heavy ion
collisions, this state can be observed in the $\Lambda N$ or
$\pi\Sigma N$ invariant mass spectrum.

\item $(\Omega\Omega)_{0^+}$: This is a deeply bound six-quark
state predicted by the chiral quark
model~\cite{di-omega,Ko-Zhang}. It has a large binding energy of
about $116$ MeV and a small root mean square distance of $0.84$ fm
between the two $\Omega$s. Because of its large
strangeness content, it is stable against strong hadronic decays
and possesses the weak decays $(\Omega\Omega)_{0^+} \to \pi^- +
\Xi^0 + \Omega^-$ and $(\Omega\Omega)_{0^+} \to \pi^0 + \Xi^- +
\Omega^-$ with a mean lifetime estimated to be about four times
longer than the free $\Omega$ lifetime of $0.822 \times 10^{-10}$
sec. Apart from these conventional decay modes, the nonmesonic
decay $(\Omega\Omega)_{0^+} \to \Xi^- +\Omega^-$ is also possible;
and the estimated lifetime of $(\Omega\Omega)_{0^+}$ for this
process is twice the free $\Omega$ lifetime. Thus, instead of
direct observation, the $(\Omega\Omega)_{0^+}$ may also be
detected in the $\Xi^-\Omega^-$ invariant mass distribution.

\item $H_c^{++}$:  This dibaryon with $J^P=0^0$ and $I=1$ is
predicted in Ref.\cite{Lee09}. It is expected to be strongly bound
as it is composed of $[ud]$, $[us]$, and $[uc]$ scalar diquarks,
and one of which has to break in order for $H_c^{++}$ to fall
apart and to decay to $p+\Xi^+$. The lifetime is expected to be
similar to that of $\Xi^+$, and the dominant hadronic weak decay
mode is expected to be $p+\Lambda K^- \pi^+ \pi^+$.

\item $\bar{D}NN$ and $BNN$: The attractive $\bar{D}N$ and $BN$
interactions would lead to bound $\bar{D}$ and $B$ mesons in the
nuclear medium as their binding energies increase with increasing
nuclear density. The $\bar{D}NN$ and $BNN$ molecular states
predicted in Ref.~\cite{Yasui:2009bz} are thus nuclei with minimum
baryon number. The quantum numbers can be $J^{P}=0^{-}$ or $1^{-}$
and $I=0$ with different types of weak decay mode. The $\bar{D}NN$
states have the decay modes $K^{+}\pi^{-}\pi^{-}+p+p$ for $0^{-}$
and $K^{+}\pi^{-}+d$ for $1^{-}$ with all charged particles in the
final state, while the $BNN$ states have the decay modes
$K^{+}\pi^{-}+\pi^{+}+p+p$ for $0^{-}$ and
$K^{+}\pi^{-}+\pi^{+}+d$ for $1^{-}$. Therefore, the experimental
observation of these decays makes it possible to determine both
the spins and parities of these hadronic molecular states.

\end{enumerate}

\section{Yields of exotic hadrons in heavy ion collisions}\label{results}

We show in this Section the expected yields of exotic hadrons
described in the previous Section from central Au+Au collisions at
$\sqrt{s_{NN}}=200$ GeV at RHIC and central Pb+Pb collisions at
$\sqrt{s_{NN}}=5.5$ TeV at LHC. They include results for all
possible structure configurations, \textit{e.g.} multiquark
hadrons and hadronic molecules, calculated from the coalescence
model in addition to those estimated from the statistical model.
These results are shown in Table~\ref{Tab:hadron}. We also give
some discussions on the obtained results.

\begin{table*}[htdp]
\caption{Exotic hadron yields in central Au+Au collisions at
$\sqrt{s_{NN}}= 200$ GeV at RHIC and in central Pb+Pb collisions
at $\sqrt{s_{NN}}=5.5$ TeV at LHC from the quark coalescence
($2q/3q/6q$ and $4q/5q/8q$) and the hadron coalescence (Mol.) as
well as from the statistical model (Stat.)}\label{Tab:hadron}
\begin{center}
\begingroup
\renewcommand{\arraystretch}{1.2}
\begin{tabular}{c|c|c|c|c|c|c|c|c}
\hline \hline & \multicolumn{4}{|c}{RHIC}
& \multicolumn{4}{|c}{LHC}\\
\cline{2-9} & $2q/3q/6q$ & $4q/5q/8q$ & Mol. & Stat. & $2q/3q/6q$ & $4q/5q/8q$ & Mol. & Stat.
\\
\hline
Mesons &&&&&&&&\\
$f_0(980)$  & 3.8, 0.73($s\sbar$) & 0.10 & 13 & 5.6 &10, 2.0 ($s\sbar$) & 0.28 & 36 & 15 \\
$a_0(980)$  & 11 & 0.31 & 40 & 17 & 31 & 0.83 & \Ex{1.1}{2} & 46 \\
$K(1460)$   & --- & 0.59 & 3.6 & 1.3 & --- & 1.6 & 9.3 & 3.2 \\
$D_{s}(2317)$   & \Ex{1.3}{-2} & \Ex{2.1}{-3} & \Ex{1.6}{-2}    & \Ex{5.6}{-2}
        & \Ex{8.7}{-2} & \Ex{1.4}{-2} & 0.10        & 0.35 \\
$T_{cc}^{1}$ $^{\dag)}$ & --- & \Ex{4.0}{-5} & \Ex{2.4}{-5} & \Ex{4.3}{-4}
            & --- & \Ex{6.6}{-4} & \Ex{4.1}{-4} & \Ex{7.1}{-3} \\
$X(3872)$ & \Ex{1.0}{-4} & \Ex{4.0}{-5} & \Ex{7.8}{-4} &
\Ex{2.9}{-4} & \Ex{1.7}{-3} & \Ex{6.6}{-4} & \Ex{1.3}{-2} & \Ex{4.7}{-3}   \\
$Z^+(4430)$$^{\ddag)}$ & --- & \Ex{1.3}{-5} & \Ex{2.0}{-5} &
\Ex{1.4}{-5} & --- & \Ex{2.1}{-4} & \Ex{3.4}{-4} & \Ex{2.4}{-4} \\
$T_{cb}^{0}$ $^{\dag)}$ & --- & \Ex{6.1}{-8} & \Ex{1.8}{-7} &
\Ex{6.9}{-7} & --- & \Ex{6.1}{-6} & \Ex{1.9}{-5} & \Ex{6.8}{-5} \\
\hline
Baryons &&&&&&&&\\
$\Lambda(1405)$ & 0.81 & 0.11 & 1.8$-$8.3 & 1.7 & 2.2 & 0.29 & 4.7$-$21 & 4.2 \\
$\Theta^+$  $^{\ddag)}$ & --- & $2.9\times 10^{-2}$ & --- & 1.0 & --- & \Ex{7.8}{-2} & --- & 2.3 \\
$\bar{K}KN$  $^{\dag)}$ & \com{---} & $1.9\times 10^{-2}$ & 1.7 & 0.28 & \com{---} & \Ex{5.2}{-2} & 4.2 & 0.67 \\
$\bar{D}N$ $^{\dag)}$ & --- & \Ex{2.9}{-3} & \Ex{4.6}{-2} & \Ex{1.0}{-2} & --- & \Ex{2.0}{-2} & 0.28 & \Ex{6.1}{-2} \\
$\bar{D}^*N$ $^{\dag)}$ & --- & \Ex{7.1}{-4} & \Ex{4.5}{-2} & \Ex{1.0}{-2} & --- & \Ex{4.7}{-3} & 0.27 & \Ex{6.2}{-2} \\
$\Theta_{cs}$ $^{\dag)}$ & --- & \Ex{5.9}{-4} & --- & \Ex{7.2}{-3} & --- & \Ex{3.9}{-3} & --- & \Ex{4.5}{-2} \\
$BN$ $^{\dag)}$ & --- & \Ex{1.9}{-5} & \Ex{8.0}{-5} & \Ex{3.9}{-5} & --- & \Ex{7.7}{-4} & \Ex{2.8}{-3} & \Ex{1.4}{-3} \\
$B^*N$ $^{\dag)}$ & --- & \Ex{5.3}{-6} & \Ex{1.2}{-4} & \Ex{6.6}{-5} & --- & \Ex{2.1}{-4} & \Ex{4.4}{-3} & \Ex{2.4}{-3} \\
\hline
Dibaryons &&&&&&&&\\
$H$ $^{\dag)}$ & \Ex{3.0}{-3} & --- & \Ex{1.6}{-2} & \Ex{1.3}{-2} & \Ex{8.2}{-3} & --- & \Ex{3.8}{-2} & \Ex{3.2}{-2} \\
$\bar{K}NN$ $^{\ddag)}$ & \Ex{5.0}{-3} & \Ex{5.1}{-4} &
0.011$-$0.24 & \Ex{1.6}{-2} & \Ex{1.3}{-2} & \Ex{1.4}{-3} & $0.026-0.54$ & \Ex{3.7}{-2} \\
$\Omega\Omega$ $^{\dag)}$ & \Ex{3.2}{-5} & --- & \Ex{1.5}{-5} & \Ex{6.4}{-5} & \Ex{8.6}{-5} & --- & \Ex{4.4}{-5} & \Ex{1.9}{-4} \\
$H_c^{++}$ $^{\dag)}$ & \Ex{3.0}{-4} & --- & \Ex{3.3}{-4} & \Ex{7.5}{-4} & \Ex{2.0}{-3} & --- & \Ex{1.9}{-3} & \Ex{4.2}{-3} \\
$\bar{D}NN$ $^{\dag)}$ & --- & \Ex{2.9}{-5} & \Ex{1.8}{-3} & \Ex{7.9}{-5} & --- & \Ex{2.0}{-4} & \Ex{9.8}{-3} & \Ex{4.2}{-4} \\
$BNN$ $^{\dag)}$ & --- & \Ex{2.3}{-7} & \Ex{1.2}{-6} & \Ex{2.4}{-7} & --- & \Ex{9.2}{-6} & \Ex{3.7}{-5} & \Ex{7.6}{-6} \\
\hline \hline
\end{tabular}
\endgroup
\end{center}
\end{table*}%

Comparisons of the yields in the $2q/3q/6q$ column to those in the
$4q/5q/8q$ column in Table~\ref{Tab:hadron} show that for most of
the hadronic states considered here, the yield from the
coalescence model for the compact multiquark state is smaller than
that for the usual quark configuration as a result of the
suppression due to the coalescence of additional quarks indicated
in Eq.~(\ref{coal-factors}). For the same state, the yield from
the coalescence model for the molecular configuration is, however,
larger than that from the statistical model prediction as seen
from comparing the yields in the Mol. column to those in the Stat.
column in Table~\ref{Tab:hadron}. This is in contrast to high
energy pp collisions, where molecular configurations with small
binding energies are hard to be produced at high
$p_T$~\cite{Bignamini:2009sk}.

\begin{figure*}
\Psfig{8cm}{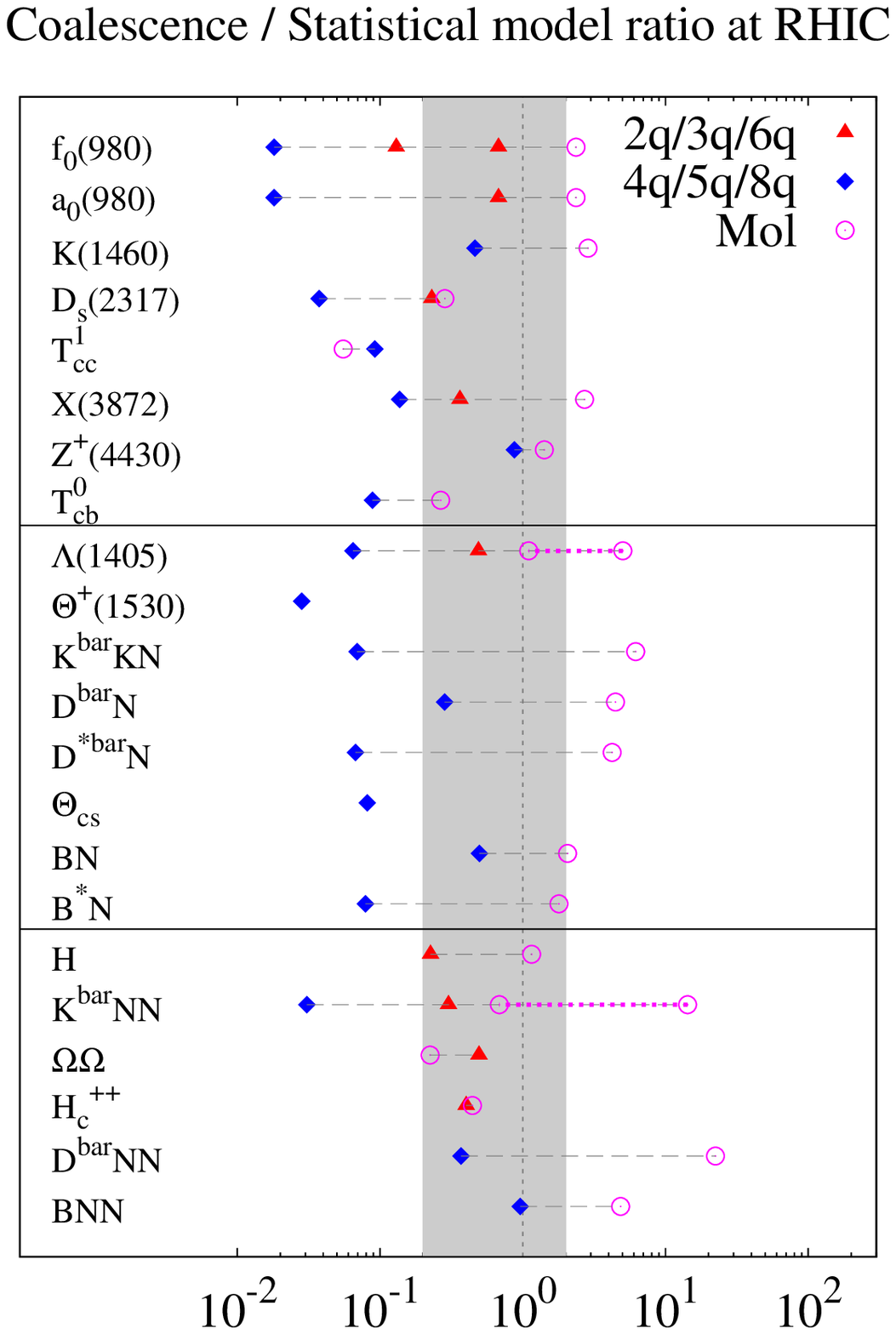}~\Psfig{8cm}{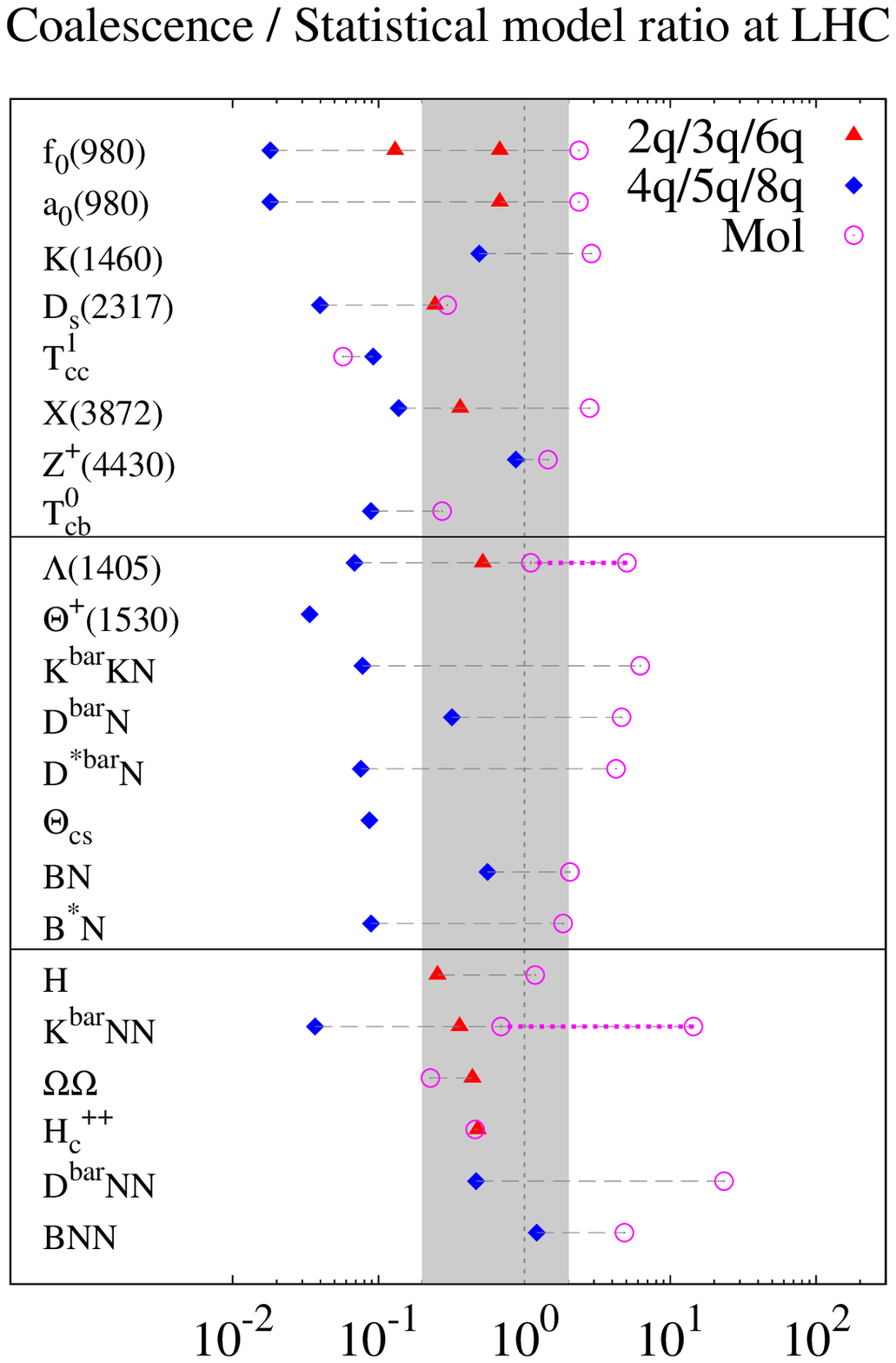}
\caption{(Color online) Ratio of the yield of an exotic hadron in the coalescence model to that of the statistical model.}
\label{Fig:List}
\end{figure*}

To see more clearly the effect of the structure of a hadron on its
production in heavy ion collisions, we show in Fig.~\ref{Fig:List}
the ratio of the coalescence model results to those from the
statistical model,
\begin{align}
R^\mathrm{CS}_h \equiv N_h^\mathrm{coal}/N_h^\mathrm{stat}
\ .
\end{align}
Generally, the ratios for the $2q$ and $3q$ configurations are
within the range of $0.2 < R_h < 2$ for normal hadrons, which is
shown by the grey zone. This also applies to exotic hadrons with
most of the $2q/3q$ configurations, as shown by the triangles. We
observe that the coalescence yield for an exotic multiquark hadron
(diamond) is smaller than those for the usual quark configuration
and from the statistical model predictions. This is consistent
with the naive expectation that the multiquark coalescence becomes
suppressed as the quark number increases. The tetraquark states
$f_0(980)$ and $a_0(980)$ are typical examples. In these hadrons,
tetraquark configurations necessarily involve strange quarks, and
they are thus more suppressed. This suppression also applies to
the $5q$ states in exotic baryons ($\Lambda(1405), \Theta^+(1530),
\bar{K}KN,$ and $\Theta_{cs}$) and the $8q$ state in the
$\bar{K}NN$.

The yields of hadronic moleculars from the hadron coalescence
(circles) depend strongly on the sizes of hadrons. For deeply
bound and compact hadronic molecules, their yields are comparable
to or smaller than the predictions of the statistical model. On
the other hand, the loosely bound extended molecules would be
formed abundantly. One typical example is $\Lambda(1405)$ that are
shown in Table~\ref{summary} for the two cases of $\omega=20.5$
and $174~\MeV$. The smaller value, corresponding to a small
binding~\cite{Jido:2003cb,Hyodo:2007jq}, gives a larger size for
the $\Lambda(1405)$. As a result, antikaons produced in heavy ion
collisions have larger probabilities of coalescing with nucleons,
resulting in more abundant production of $\Lambda(1405)$. On the
other hand, for the larger $\omega$ value, which corresponds to
the case that the pole position of the S-matrix for the two-hadron
interaction is around $1405~\MeV$, the $\Lambda(1405)$ can be
regarded as a deeply bound state and thus has a smaller size, and
hence its yield becomes smaller.

\section{Discussions}\label{discussion}

Our results based on the coalescence model for hadron production
in relativistic heavy ion collisions have indicated that their
yields are strongly dependent on their structures. Therefore,
measuring the yields of exotic hadrons allows us to infer the
internal configuration of exotic hadrons \cite{Cho:2010db,
Ohnishi:2011nq}. For example, we have mentioned in Sec.
\ref{results} A that as possible configurations for $f_0(980)$,
quark-antiquark pairs ($\sim s \bar{s}$, $u \bar{u}$, and $d
\bar{d}$), a tetraquark state, and a $K \bar{K}$ hadronic molecule
have been proposed. To confirm its structure, we refer to
preliminary data from the STAR Collaboration for the production
yield ratios of $f_0(980)$, $\pi$, and
$\rho^0$\cite{Fachini:2002yj}. From these results we find that the
measured yield of $f_0(980)$ is close to 8, which means that it is
more probable for the $f_0(980)$ to be produced as a hadronic
molecule state than a tetraquark state (See the order of magnitude
difference between the yield in the $4q/5q/8q$ column and that in
the Mol. column in Table~\ref{Tab:hadron}). Therefore, we conclude
that the STAR data seem to rule out a dominant tetraquark
configuration for the $f_0(980)$. Further experimental efforts to
reduce the error bar are thus highly desirable.

For some exotic hadrons, our results show that the yields are
similar for the hadronic and the molecular configuration, despite
the difference in the coalescence temperatures $T_C$ and $T_F$.
This can be attributed to the larger size of the molecular
configuration. Assuming other factors are similar, the $s$-wave
factors involved in the coalescence at $T_F$ are similar to those
at $T_C$ as long as the relevant molecular size is related to the
hadron size  as $\sigma_C=(V_C/V_F)^{1/3} \sigma_F$ as can be
inferred from Eq.~(\ref{coal-factors}) after neglecting the
temperature dependence in the denominator. If we additionally
assume that the volume scales as $V \propto T^{-3}$, we find that
the condition for the molecular coalescence to be similar to
two-quark coalescence is that the molecular size scales as
$\sigma_F =\sigma_C T_C/T_F$, which is more or less satisfied by
some exotic hadrons considered here, such as the $D_s(2317)$.

\begin{figure*}
\Psfig{14cm}{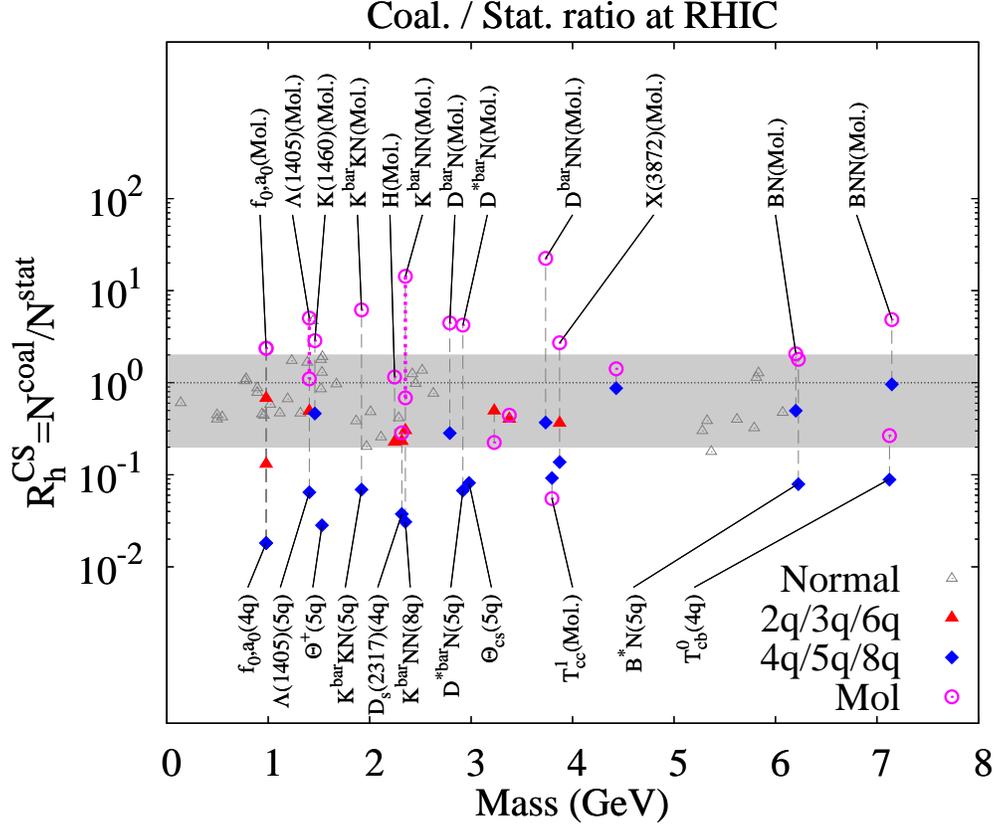} \caption{(Color online) Ratio of
hadron yields at RHIC in the coalescence model to those in the
statistical model as a function of mass.} \label{Fig:Mass}
\end{figure*}

Our study also shows the interesting result that the ratio
$R^\mathrm{CS}_h$ of the yield in the coalescence model to that in
the statistical model is almost the same at RHIC and LHC. This
similarity comes from the universal feature of the QCD phase
transition; the common critical temperature and the common volume
ratio $V_C/V_H$. In the non-relativistic approximation shown in
Eq.~(\ref{Eq:StatNR}), it is possible to rewrite the statistical
model yield in the coalescence-like form,
\begin{align}
N_h^\mathrm{stat}=
    \frac{g_h V_H (m_h T_H)^{3/2}}{(2\pi)^{3/2}}
    e^{\mathrm{B}/T_H}
    \prod_i \frac{N_{i,H}(2\pi)^{3/2}}{g_i V_H (m_i T_H)^{3/2}}
\ ,
\label{Eq:StatProd}
\end{align}
where we consider the hadron $h$ to be composed of several
constituents, $\mathrm{B}=\sum_i m_i - m_h$ is the binding energy,
and $N_{i,H}$ represents the yield of the $i$-th constituent at
the volume $V_H$. This relation holds since the fugacity of a
particle is equal to the product of the constituent fugacities,
and the particle fugacity is related to the yield according to
\begin{align}
\gamma_i = \frac{N_{i,H}}{g_h V_H}\,e^{m_h/T_H}\left(\frac{2\pi}{m_h T_H}\right)^{3/2}
\ .
\end{align}

By using Eqs.~(\ref{Eq:StatProd}) and (\ref{nquark}), the
ratio $R^\mathrm{CS}_h$
is found to be approximately given as
\begin{align}
R^\mathrm{CS}_h =&
    e^{-\mathrm{B}/T_H}
    \frac{M^{3/2}(2T_H/\omega)^{3(n-1)/2}(2T_c/\omega)^L}
    {m_h^{3/2}(1+2T_c/\omega)^{n-1+L}}
    \nonumber\\
    &\times
    \frac{V_C}{V_H}
    \prod_i \frac{N_i/V_C}{N_{i,H}/V_H}
    \prod_{j=1}^{n-1} \frac{(2l_j)!!}{(2l_j+1)!!}
\ . \label{Eq:RatioNR}
\end{align}
Since $T_C$ and $T_H$ are the same at small baryon chemical
potential, the particle density $N_i/V_C, N_{i,H}/V_H$ and the
volume ratio $V_C/V_H$ would be common at RHIC and LHC. As a
result, the ratio of the production yield of hadrons in the
coalescence model to that in the statistical model becomes the
same at RHIC and LHC.

In Fig.~\ref{Fig:Mass}, we show the hadron mass dependence of the
ratio $R_h^\mathrm{CS}$ in heavy ion collisions at RHIC. The
results are similar in heavy ion collisions at LHC as discussed in
the above. It is seen that the effect of the internal structures
of exotic hadrons on $R_h^\mathrm{CS}$ is particularly large for
light exotic hadrons. We also show the ratio for normal hadrons.
Here we have included the resonance decay contribution to
pseudoscalar mesons such as $\rho\to2\pi$ in evaluating the ratio.
This estimate would be closer to the observational condition. As
already mentioned, the ratios for normal hadrons are found to be
in the range of $0.2-2$.

It should be noted that the coalescence model may overestimate the
yield of very loosely bound molecules. Since the average hadron
distance is around 2 fm and the temperature is $T_F=125~\MeV$ at
the thermal freeze-out, the loosely bound and spatially extended
hadrons would dissociate easily through the final-state
interactions with other hadrons. For example, the deuteron yield
is calculated to be around 1.4 per unit rapidity, which is larger
than the statistical model prediction ($\sim 0.3$). Experimental
data seems to be consistent with the statistical model result,
suggesting the possibility of later coalescence of loosely bound
particles with a few MeV binding energies~\cite{oh07}.

\section{Summary}
\label{sec:summary}

In this article, we have proposed a new approach of studying
exotic hadrons in relativistic heavy ion collisions at RHIC and
LHC. We have considered the yields of proposed exotic hadrons;
$f_0(980)$, $a_0(980)$, $K(1460)$, $D_s(2317)$, $T_{cc}^1$,
$X(3872)$, $Z^+(4430)$, and $T_{cb}^0$ for exotic mesons,
$\Lambda(1405)$, $\Theta^+(1530)$, $\bar{K}KN$, $\bar{D}N$,
$\bar{D^*}N$, $\Theta_{cs}$, $BN$ and $\bar{B^*}N$ for exotic
baryons, $H$, $\bar{K}NN$, $\Omega\Omega$, $H_c^{++}$, $\bar{D}NN$
and $BNN$ for exotic dibaryons. To obtain the yields of these
exotic multiquark hadrons or hadronic molecular states, we have
used the coalescence model based on either the quark degrees of
freedom or the hadronic degrees of freedom.

Our results indicate that the yields of many exotic hadrons are
large enough to be measurable in experiments. In particular, heavy
exotic hadrons containing charm and bottom quarks as well as
strange quarks can be possibly observed at RHIC and especially at
LHC. Therefore, relativistic heavy ion collisions will provide a
good opportunity to search for exotic hadrons. In particular, it
may lead to the first observation of new exotic hadrons. Also, we
have found that the structure of light exotic hadrons has a
significant effect on their yields in heavy ion collisions. For a
hadron of normal quark structure, its production yield relative to
the statistical model prediction $R_h=N_h/N_h^\mathrm{stat}$ is
found in the range of $0.2-2$. The yield ratio is smaller ($R_h <
0.2$) if a hadron has a compact multiquark configuration. For a
hadron of molecular configuration with an extended size, its yield
is, on the other hand, larger than the normal values ($R_h > 2$).
Therefore, the ratios of measured yields from experiments on heavy
ion collisions to those predicted by the statistical model
provides a new method to discriminate the different pictures for
the structures of exotic hadrons. The study of exotic hadrons in
relativistic heavy ion collisions thus will help answer
longstanding problems on the existence and structure of exotic
hadrons.

\section*{ACKNOWLEDGEMENTS}

This work was supported in part by the Yukawa International
Program for Quark-Hadron Sciences at Yukawa Institute for
Theoretical Physics, Kyoto University, the Korean Ministry of
Education through the BK21 Program and KRF-2006-C00011, the
Grant-in-Aid for Scientific Research (Nos. 21840026 and 22105507
and 22-3389), the Grant-in-Aid for Scientific Research on Priority
Areas ``Elucidation of New Hadrons with a Variety of Flavors''
from MEXT (Nos. 21105006 and 22105514), the Grant-in-Aid for the
global COE program "The Next Generation of Physics, Spun and from
Universality and Emergence" from MEXT, the Brazilian Research
Council (CNPq) and the S\~{a}o Paulo State Research Foundation
(Fapesp), the U.S. National Science Foundation under Grants No.
PHY-0758115 and No. PHY-1068572, and the Welch Foundation under
Grant No. A-1358. We thank the useful discussions with other
participants during the YIPQS International Workshop on "Exotics
from Heavy Ion Collisions" when this work was started. T.S.
acknowledges the support by the Grand-in-Aid for JSPS fellows.
T.H. further thanks the support from the Global Center of
Excellence Program by MEXT, Japan through the Nanoscience and
Quantum Physics Project of the Tokyo Institute of Technology.

\appendix

\section{d-wave Wigner function}
\label{App:A}

In this Appendix, we extend the calculation shown in
Ref.~\cite{baltz&dover} to construct the $d$-wave Wigner function
from the harmonic oscillator wave functions. For this, we need the
wave function of the second excited state, in addition to those of
the ground and the first excited state, as the basis functions for
the Wigner function,

\begin{align}
\phi_0(x)=&\Big(\frac{1}{b\sqrt{\pi}}\Big)^{\frac{1}{2}}e^{
-\frac{x^2}{2b^2}} \nonumber \\
\phi_1(x)=&\Big(\frac{2}{b\sqrt{\pi}}\Big)^{\frac{1}{2}}\frac{
x}{b}e^{-\frac{x^2}{2b^2}} \nonumber \\
\phi_2(x)=&\Big(\frac{1}{2b\sqrt{\pi}}\Big)^{\frac{1}{2}}\Big(
2\frac{x^2}{b^2}-1\Big)e^{-\frac{x^2}{2b^2}}
\end{align}
where $b^2=\hbar/(m\omega)$. The $M$-state wave functions of the
$L=2$ state in Cartesian coordinates, are then

\begin{align}
\phi_{M=2}^{L=2}(\vec r)=&\frac{1}{2}\bigg[\Big(\phi_2(x)
\phi_0(y)-\phi_0(x)\phi_2(y)\Big) \nonumber \\
&+i\sqrt{2}\phi_1(x)\phi_1(y)\bigg] \phi_0(z) \nonumber \\
\phi_{M=1}^{L=2}(\vec r)
=&\frac{1}{\sqrt{2}}\Big(\phi_1(x)
\phi_0(y)+i\phi_0(x)\phi_1(y)\Big)\phi_1(z) \nonumber \\
\phi_{M=0}^{L=2}(\vec r)=&\frac{1}{\sqrt{6}}\Big(2\phi_0(x)
\phi_0(y)\phi_2(z) \nonumber \\
&-\phi_2(x)\phi_0(y)\phi_0(z)-\phi_0(x)\phi_2(y)
\phi_0(z)\Big) \nonumber \\
\phi_{M=-1}^{L=2}(\vec r)=&\phi_{M=1}^{L=2}(\vec r)^*\nonumber\\
\phi_{M=-2}^{L=2}(\vec r)=&\phi_{M=2}^{L=2}(\vec r)^*
\end{align}
From the density matrix of the $L=2$ state, which is defined by
averaging over the $M$ states,

\begin{align}
\rho^{L=2}(\vec r,\vec r')
=&\frac{1}{5}\Big(
\phi_{M=2}^{L=2}(\vec r)\phi_{M=2}^{L=2}(\vec r')^*
+\phi_{M=1}^{L=2}(\vec r)\phi_{M=1}^{L=2}(\vec r')^*
\nonumber \\
+& \phi_{M=0}^{L=2}(\vec r)\phi_{M=0}^{L=2}(\vec r')^*
+\phi_{M=-1}^{ L=2}(\vec r)\phi_{M=-1}^{L=2}(\vec r')^*
\nonumber \\
+&\phi_{M=-2}^{L=2}(\vec r)\phi_{M=-2}^{L=2}(\vec r')^*\Big),
\end{align}
we obtain the $M$-averaged $L=2$ harmonic oscillator Wigner
function,

\begin{align}
f^{L=2}(\vec r, \vec k)=&\int \rho^{L=2}(\vec
r+\frac{\vec\eta}{ 2},\vec r-\frac{\vec\eta}{2})e^{i\vec
k\cdot\vec\eta}d\vec\eta
\nonumber \\
=&\frac{16}{30}\bigg[4\frac{r^4}{b^4}-20\frac{r^2}{b^2}
+15-20b^2k^2+4b^4k^4 \nonumber \\
&+16r^2k^2-8(\vec r\cdot\vec k)^2\bigg]e^{
-\frac{r^2}{b^2}-b^2k^2}. \label{dWigner}
\end{align}
By replacing $b$ with $\sigma(=1/\sqrt{\mu\omega})$, we obtain the
$d$-wave coalescence factor using Eqs. (\ref{Eq:Coal}) and
(\ref{dWigner}) as

\begin{equation}
\frac{\int d^3yd^2k\tilde{f}(\vec k)f^W_d(\vec y,\vec k)}{\int
d^3yd^2k\tilde{f}(\vec k)}=\frac{(4\pi\sigma^2)^{\frac{3}{2}}}{
V(1+2\mu T\sigma^2)^3} \frac{8}{15}(2\mu T\sigma^2)^2\ .
\end{equation}

\section{Coalescence factor for the general angular momentum $l$}
\label{App:B}

We derive in this Appendix the coalescence factor for a general
angular momentum $l$-state given by Eq.~(\ref{cf}). For
constituents that are uniformly distributed in space, we can
integrate the Wigner functions over space to obtain the momentum
distribution of the constituents in the hadron. For the harmonic
oscillator wave function of the lowest energy state with a given
$l$, it is given by

\begin{equation}
P_l (k) = (4\pi \sigma^2)^{3/2} \frac{(2\sigma^2k^2)^l}{(2l+1)!!}
e^{-\sigma^2 k^2}
\end{equation}
with the normalization $\int P_l (k) d^3 k/(2\pi)^3 = 1$. The
coalescence factor is then given by

\begin{align}
F(\sigma, \mu, l, T) =& \int d^2 k P_l (k) e^{-\frac{k^2}{2\mu T}} /
\int d^2 k e^{-\frac{k^2}{2\mu T}}  \nonumber \\
=&\frac{(4\pi \sigma^2)^{3/2}}{1+2\mu T \sigma^2} \frac{(2 l)
!!}{(2l+1)!!} \left[\frac{2\mu T \sigma^2}{1+2\mu T
\sigma^2}\right]^l \ .
\end{align}

\end{document}